\documentclass[a4paper,titlepage,12pt]{article}
\pdfoutput=1
\usepackage{bm}
\usepackage[utf8]{inputenc}
\usepackage{graphicx}
\usepackage{amsbsy}
\usepackage{amsmath}
\usepackage{amsfonts}
\usepackage{amsthm}
\usepackage{mciteplus}
\usepackage{listings}
\usepackage{braket}
\usepackage{float}
\usepackage{url}
\usepackage{bbold}
\usepackage{pstricks}
\usepackage{hyperref}  
\usepackage{footmisc}
\usepackage{multirow}
\usepackage{blindtext}
\usepackage{titlesec}
\usepackage{epigraph}
\usepackage{CJKutf8}
\usepackage[english]{babel}
\usepackage[OT2,T1]{fontenc}
\usepackage[font={small,it}]{caption}

\usepackage{geometry}

 \usepackage[sep=2pt, offset=1em]{simpler-wick}

\graphicspath{ {./images/} }

\numberwithin{equation}{section}

\usepackage{amssymb}
\usepackage{tikz-cd}
\usepackage{tikz}
\usepackage{mathrsfs}
\usepackage{footmisc}

\hypersetup{
    colorlinks,
    citecolor=blue,
    filecolor=blue,
    linkcolor=blue,
    urlcolor=blue
}

\def\be{\begin{equation}}
\def\ee{\end{equation}}

\def\mg{\mathfrak{g}}

\def\tr{\text{tr}}

\def\tx{\text}

\theoremstyle{definition}

\definecolor{darkgreen}{rgb}{0.0, 0.5, 0.0}

\newcommand{\RNum}[1]{\uppercase\expandafter{\romannumeral #1\relax}}

\title{Factorizing Chern-Simons Theory via Quantum Group}
\author{Thomas G. Mertens \\
Qi-Feng Wu (吴奇峰)}

\begin{document}
\begin{CJK*}{UTF8}{gbsn}


\begin{titlepage}

\setcounter{page}{1} \baselineskip=15.5pt \thispagestyle{empty}

\vfil

${}$
\vspace{1cm}

\begin{center}

\def\thefootnote{\fnsymbol{footnote}}
 
\begin{center}
\hspace{-0.5cm}
{\Large \bf Minimal Factorization of Chern-Simons Theory} \\
\vspace{0.2cm}
{\large \bf - Gravitational Anyonic Edge Modes -}
\end{center}

~\\[1cm]
{Thomas G. Mertens,\footnote{{\protect\path{thomas.mertens@ugent.be}}} Qi-Feng Wu (吴奇峰),\footnote{{\protect\path{qifeng.wu@ugent.be}}}}
\\[0.3cm]
{\normalsize { \sl Department of Physics and Astronomy
\\[1.0mm]
Ghent University, Krijgslaan, 281-S9, 9000 Gent, Belgium}}\\[3mm]

\end{center}

 \vspace{0.2cm}
 
{\small  \noindent 
\begin{center} 
\textbf{Abstract}
\end{center} 
One approach to analyzing entanglement in a gauge theory is embedding it into a factorized theory with edge modes on the entangling boundary. For topological quantum field theories (TQFT), this naturally leads to factorizing a TQFT by adding local edge modes associated with the corresponding CFT. In this work, we instead construct a minimal set of edge modes compatible with the topological invariance of Chern-Simons theory. This leads us to propose a minimal factorization map. These minimal edge modes can be interpreted as the degrees of freedom of a particle on a quantum group. Of particular interest is three-dimensional gravity as a Chern-Simons theory with gauge group SL$(2,\mathbb{R}) \times$ SL$(2,\mathbb{R})$. Our minimal factorization proposal uniquely gives rise to quantum group edge modes factorizing the bulk state space of 3d gravity. This agrees with earlier proposals that relate the Bekenstein-Hawking entropy in 3d gravity to topological entanglement entropy.
}

 \vspace{0.3cm}
\vfil
\begin{flushleft}
\today
\end{flushleft}

\end{titlepage}

\tableofcontents

\setcounter{footnote}{0}

\section{Introduction}
As a universal phenomenon, entanglement plays a crucial role in understanding the emergence of spacetime \cite{Ryu:2006bv,VanRaamsdonk:2010pw}. In a quantum theory, characterizing entanglement requires a factorization of the state space, e.g.
\begin{equation}
    \mathcal{H} = \mathcal{H}_A \otimes \mathcal{H}_{\Bar{A}}.
\end{equation}
In gauge theories, factorizing the state space becomes subtle due to the presence of nonlocal degrees of freedom. One resolution is to embed it into a factorized state space by adding degrees of freedom, called edge modes, on the entangling boundary \cite{Donnelly:2011hn}. 

This approach is ambiguous since the embedding is not unique. An intuitive way to see this is as follows. Suppose a state space $\mathcal{H}$ is embedded into a factorized space $\mathcal{H}_A \otimes \mathcal{H}_{\Bar{A}}$ with the embedding map
\begin{alignat}{4}
\label{eq: general factorization}
F:\; && \mathcal{H} && \;\hookrightarrow\;     & \mathcal{H}_A \otimes \mathcal{H}_{\Bar{A}}.
\end{alignat}
We can attach one qubit degree of freedom to $\mathcal{H}_A$ and $\mathcal{H}_{\Bar{A}}$ respectively such that the embedding map is modified as
\begin{alignat}{4}
F':\; && \mathcal{H} && \;\hookrightarrow\;     & (\mathcal{H}_A \otimes \mathbb{C}^2) \otimes 
(\mathbb{C}^2 \otimes  \mathcal{H}_{\Bar{A}}).
\end{alignat}
If the attached pair of qubits is in a Bell state, then the entanglement entropy will increase by one unit. In this way, one can arbitrarily enlarge the extended state space and increase the entanglement. From this perspective, quantities calculated in the extended state space formalism are not necessarily intrinsic or physical. One can remove the additional qubit state spaces and still be able to reproduce the original non-factorized state space. However, one cannot arbitrarily reduce the extended state space. This naturally leads to the question:
\begin{quotation}
\centering 
\textit{What is the minimal extension that factorizes a gauge theory?}
\end{quotation}

Factorization of the state space in gauge theories has a long history, see e.g.
\cite{Buividovich:2008gq,Casini:2013rba,Donnelly:2014gva,Lin:2018bud,Ghosh:2015iwa} for a general framework,  \cite{Donnelly:2015hxa,Blommaert:2018rsf,Blommaert:2018oue,Geiller:2019bti} for concrete examples in Maxwell and Yang-Mills theory. A crucial role is played by local degrees of freedom that live at the entangling surface, called edge modes, that facilitate a factorization of the model. These degrees of freedom are fictitious according to an observer who has access to the whole space $A \cup \Bar{A}$ (``two-sided observer''), but are an intrinsic part of the state space of an observer who only has access to a portion of the state space, $A$ or $\Bar{A}$ (``one-sided observer''). The introduction of the edge modes can be formalized by defining a so-called factorization map \cite{Donnelly:2016jet,Donnelly:2018ppr,Donnelly:2020teo,Jiang:2020cqo}.Topological gauge theories, such as 3d Chern-Simons field theory and BF gauge theories, are also part of this class of theories. In particular, it is a long-standing belief that factorizing 3d Chern-Simons gauge theory leads to a WZNW model describing the edge degrees of freedom at the entangling surface \cite{Wen:2016snr,Wong:2017pdm,Geiller:2017xad,Fliss:2020cos,Klinger:2023qna}.
However, Chern-Simons theory does not just have gauge invariance, but also has topological invariance, which further reduces degrees of freedom. This implies that adding Kac-Moody edge modes is not the minimal extension.

In this work, we will illustrate factorization at the phase space level. What we are seeking is a surjective map dual to the factorization map \eqref{eq: general factorization},
\begin{equation}
\label{eq: general gluing morphism}
    G:\mathcal{P}_A \times \mathcal{P}_{\bar{A}} \twoheadrightarrow \mathcal{P}.
\end{equation}
$\mathcal{P}_A$, $\mathcal{P}_{\bar{A}}$, and $\mathcal{P}$ are phase spaces of $A$, $\bar{A}$, and the full system respectively. The gluing map \eqref{eq: general gluing morphism} must preserve the Poisson algebra, i.e. is a Poisson map, to ensure that the dual factorization map \eqref{eq: general factorization} preserves the operator algebra. The gluing map \eqref{eq: general gluing morphism} must be surjective to ensure that the full system can be recovered. Finding the gluing map \eqref{eq: general gluing morphism} is equivalent to taking the "square root" of the Poisson algebra on $\mathcal{P}$.\footnote{A Poisson algebra is the algebra of observables in a phase space equipped with a Poisson bracket.} Taking the square root in a theory often leads to a nontrivial extension. The Pythagorean school discovered irrational numbers by taking the square root of rational numbers. Dirac discovered the spinor by taking the square root of the Klein-Gordon equation. In the same spirit, we will take the square root of the Poisson algebra of Chern-Simons theory on a manifold with boundaries, which leads to the Poisson algebra of a chiral WZNW model coupled with a particle on a quantum group. Schematically,\footnote{The name "$\sqrt{\tx{Chern-Simons}}$" is inspired by the bulk-boundary relation in (2+1) dimensional topological orders where the category of the boundary theory is the "square root" of the modular tensor category of the bulk theory~\cite{kong2018drinfeld, kong2024enriched}.}
\begin{equation}
\label{eq: sqrt CS = chi WZNW times PQG}
    \sqrt{\tx{Chern-Simons}} = \chi \tx{WZNW} \times \tx{Particle-on-Quantum-Group}.
\end{equation}

Our main result is constructing a minimal factorization map, with associated edge modes. These edge degrees of freedom transform as representations of the corresponding quantum group, which we claim is natural for non-compact gauge groups. 
Our endeavors were motivated by recent investigations in 3d gravity, where factorization based on the WZNW model at the entangling surface was found to be contradictory with holography \cite{Mertens:2022ujr}, corroborating an older proposal of \cite{McGough:2013gka}. Related investigations into the factorization and symmetry properties of 3d gravity from various perspectives were performed in \cite{Delcamp:2016eya, Akers:2024wab, Dupuis:2020ndx,Chua:2023ios, Chen:2024unp}.\footnote{For 2d JT gravity, a similar investigation was performed in \cite{Blommaert:2018iqz}. And for Liouville gravity and DSSYK, recent investigations include \cite{Mertens:2022aou,Belaey:2025ijg}, where quantum group edge states were constructed in a 2d gravity bulk.}

More generally, symmetries are at the heart of essentially all exactly solvable physical models, with quantum group structures taking a prominent place in this list. 
In particular quantum group structures were studied extensively in Chern-Simons theory \cite{Witten:1989wf, Guadagnini:1989th,Guadagnini:1989tj}, in rational CFT \cite{Alvarez-Gaume:1988bek,Alvarez-Gaume:1988izd,Alvarez-Gaume:1989blj} and in the chiral WZNW CFT \cite{Gawedzki:1990jc,Alekseev:1990vr,Chu:1991pn,Falceto:1992bf,Gaberdiel:1994vv}. In topological field theories and dual CFTs, quantum groups appear naturally in terms of fusion and braiding data of Wilson lines and conformal blocks. These two dual descriptions match since the same modular tensor category governs both models. The appearance of quantum groups here is solely in terms of the representation labels themselves, and one does not have access to the different states within a representation; the quantum group symmetry is a \emph{hidden} symmetry \cite{Slingerland:2001ea}. In chiral WZNW models, the quantum group symmetry emerges in a perhaps more sophisticated way \cite{Gawedzki:1990jc,Falceto:1992bf}: when attempting to make sense of a chiral WZNW model by splitting a non-chiral WZNW model in two pieces, the closedness of the chiral symplectic structure requires deformation compared to the naive one, immediately leading towards the quantum group governing the right multiplication gauge symmetry of the left-movers and vice versa. In this set-up, the quantum group symmetry is more manifest; it is ``unhidden'' and the different states within a representation are accessible in the Hilbert space of the model.
In this work, we explore a related way in which the quantum group structure emerges in Chern-Simons theory, when minimally factorizing the state space of the model across an entangling surface.

One can get inspiration for how to do this from the following observation. The Chern-Simons path integral (with Lie algebra $\mathfrak{g}$) on a ``hollow'' torus $S^1 \times I \times S^1$, with conformal boundaries labeled by conformal structure moduli $\tau_1$ and $\tau_2$, is given by
$Z(\tau_1,\tau_2) = \sum_h \bar{\chi}_h(\tau_1)\chi_h(\tau_2)$, where one sums over all lowest weight (or primary) labels $h$ (all integrable representation of $\hat{\mathfrak{g}}$). This expression is identical to that of a non-chiral WZNW CFT, based on the same Lie algebra $\mathfrak{g}$, but now with each chiral WZNW sector geometrically localized on one torus wall of the 3d geometry. As mentioned above, considerable attention in the past has gone into factorizing the non-chiral WZNW models into its chiral components \cite{Gawedzki:1990jc,Falceto:1992bf}. We see here that this is essentially the same problem as geometrically factorizing Chern-Simons theory across an entangling surface into a left sector (living on one torus boundary) and a right sector (living on the other torus boundary). We will build on this observation in the appendices, but our main text is written as an independent first principles construction. \\

\textbf{Comparison with literature} \\
Our resulting Poisson bracket relations of ``one-sided'' Wilson lines in subsection \ref{s:sqrch} are closely related to (and consistent with) those of Fock and Rosly \cite{Fock:1998nu} and their subsequent ``combinatorial quantization'' \cite{Alekseev:1994pa,Alekseev:1994au}. This technique was applied to 3d flat gravity in e.g. \cite{Meusburger:2003ta,Ballesteros:2015twa}. A key difference is that theses works \emph{postulate} an $r$-matrix associated to each puncture on a discretized spatial manifold, and then show that this is mathematically consistent, and leads to a structurally solid quantization procedure. We instead will start from first principles with just the Lagrangian and symplectic (or Poisson) structure, and derive the presence of an $r$-matrix. A second difference, is that we will go beyond the Poisson brackets of Wilson lines (or loops) and construct a phase space algebra localized on the entangling surface. \\
The works \cite{Meusburger:2008bs,Freidel:2021ajp} study the emergence of the Drinfel'd double of a classical Lie group within 3d flat gravity, as a symmetry structure of the Hilbert space on a discretized spatial manifold. We study Chern-Simons theory for \emph{semi-simple} Lie groups, for which we end up with the Heisenberg double of a \emph{quantum} group, of relevance to $\Lambda<0$ and $\Lambda >0$ 3d Lorentzian gravity. \\

In the works \cite{Delcamp:2016yix,Delcamp:2016eya}, and more recently \cite{Akers:2024wab}, a factorization map similar to ours is studied. 
Their construction is geared towards finding a different factorization map (using the ``magnetic center'' instead of the usual electric center construction) than the conventional one introduced by Donnelly \cite{Donnelly:2014gva}. Their technical constructions are in the language of finite groups in a lattice context. We instead will use Lie groups (both compact and non-compact) directly in the continuum. Our construction and methodology are quite distinct. \\

The recent work \cite{Dupuis:2020ndx} (see also the earlier \cite{Bonzom:2014wva,Dupuis:2017otn}) studied the origins of the quantum group symmetry in 3d gravity, similarly to us motivated by earlier constructions in the literature being somewhat ad hoc and mathematical in nature. Their construction uses a BF formulation, and finds the relevant structure at the level of a boundary term in the action. They land on similar algebraic structures to us, but their language and methodology is quite different.
More broadly, the appearance of quantum groups in 3d Chern-Simons theory \cite{Witten:1989wf, Guadagnini:1989th,Guadagnini:1989tj,Moore:1989ni,Reshetikhin:1991tc} has mostly focused on their categorical structure (fusion and braiding rules). The actual quantum algebra and quantum group structures have always been somewhat more mysterious on how to make these explicit by elementary means of the Lagrangian and symplectic structure of the model. What we will achieve in this work, is a streamlined first principles derivation of the quantum group phase space algebra of the edge degrees of freedom at the entangling surface of 3d Chern-Simons theory. \\

In order to sharpen what we aim for, we first posit three key ingredients that any edge sector description should contain. We will explicitly describe and construct these ingredients in \textbf{sections \ref{s:prop1}}, \textbf{\ref{s:prop2}} and \textbf{\ref{s:prop3}} respectively. \textbf{Section \ref{s:quant}} describes how one quantizes the uncovered structure.
\begin{itemize}
\item \textbf{Nonlinear edge charge algebra} \\
Within a path integral, any gauge theory on a manifold $\mathcal{M}$ can be split into pieces by gluing together the left- and right gauge field at the boundary $\partial \mathcal{M}$ \cite{Blommaert:2018oue,Geiller:2019bti}. This can be implemented by a vector-valued Lagrange multiplier $J^\mu$:
\begin{align}
\label{eq:edgestart}
Z=\int \mathcal{D}A_\mu e^{iS[A]} &=  \int \mathcal{D}A_{L\mu} \mathcal{D}A_{R\mu} \delta(A_L\vert_{\partial \mathcal{M}}-A_R\vert_{\partial \mathcal{M}})e^{iS_L[A_L] + iS_R[A_R]} \\
&= \int \mathcal{D}A_{L\mu} \mathcal{D}A_{R\mu} \mathcal{D}J^\mu e^{iS_L[A_L] + iS_R[A_R]+ i\oint _{\partial \mathcal{M}}\text{Tr}(J^\mu(A_{L\mu}-A_{R_\mu}))}.
\end{align}
In terms of the one-sided (say left) path integral, with fixed surface current $J^\mu$,
\begin{equation}
Z_L(J^\mu) = \int \mathcal{D}A_{L\mu}  e^{iS_L[A_L] + i\oint _{\partial \mathcal{M}}\text{Tr}(J^\mu A_{L\mu})},
\end{equation}
regluing occurs by integrating over the common surface current density:
\begin{equation}
Z = \int \mathcal{D}J^\mu Z_L(J^\mu) Z_R(J^\mu).
\end{equation}
One can now describe the fully one-sided theory as
\begin{equation}
\label{eq:fullone}
Z_L \equiv \int \mathcal{D}J^\mu Z_L(J^\mu)=\int \mathcal{D}A_{L\mu} \mathcal{D}J^\mu e^{iS_L[A_L] + i\oint _{\partial \mathcal{M}}\text{Tr}(J^\mu A_{L\mu})},
\end{equation}
where we path integrate also over the surface current density on one side, and we consider large gauge transformations on $\partial \mathcal{M}$ as physical. This means we view $J^\mu$ here as additional degrees of freedom of the single-sided model. This in general non-abelian surface current $J^\mu(\vec{x},t)$ lives solely on the boundary: $(\vec{x},t) \in \partial \mathcal{M}$ with $\vec{x}$ the spatial coordinate on the edge. At this point, we have to specify the type of boundary. We are interested in the case where the spacetime boundary $\partial \mathcal{M}$ is a null surface (a horizon). In this case, one can show that a natural set of boundary degrees of freedom are just the $\mu=0$ components of the current densities $J^\mu$ \cite{Donnelly:2015hxa,Donnelly:2016auv,Blommaert:2018oue}. For $d$-dimensional Yang-Mills theory, the above formulation then leads directly to the surface charge density algebra:
\begin{equation}
\label{eq:curalg}
\{J^0_a(\vec{x}),J^0_b(\vec{y})\} = \delta(\vec{x}-\vec{y})f_{ab}{}^c J_c^0(\vec{x}).
\end{equation}
The canonically conjugate degrees of freedom of $J^\mu(\vec{x})$ can be shown to be the boundary group elements $g(x)$, viewed as matrix-valued fields. This follows from the Lagrangian source term $\text{Tr}(J^\mu A_{\mu})$, where one considers a pure gauge field $A_\mu = \partial_\mu g g^{-1}$ \cite{Blommaert:2018oue,Geiller:2019bti}, which represents a large gauge transformation on the boundary that has become physical there. The above Poisson algebra \eqref{eq:curalg} is hence extended by its action on the conjugate variables as \cite{Donnelly:2016auv}
\begin{align}
\{J_a^0(\vec{x}),g(\vec{y})\} &= \delta(\vec{x}-\vec{y})g(\vec{y})T_a, \\
\{g(\vec{x}),g(\vec{y})\} &= 0.
\end{align}
Integrating these relations over a spatial slice of $\partial \mathcal{M}$, and introducing the zero-mode variables $Q_a = \int d\vec{x} \, J^0_a(\vec{x})$ and $h = \int d\vec{x} \, g(\vec{x})$, this leads to the final Poisson algebra 
\begin{align}
\label{eq:1chalg}
\{Q_a,Q_b\} &= f_{ab}{}^c Q_c, \\
\label{eq:3chalg}
\{Q_a,h\} &= h T_a, \\
\label{eq:2chalg}
\{h,h\} &= 0.
\end{align}
Within the canonical quantization framework, the charge algebra \eqref{eq:1chalg} is quantized and results in a state space spanned by $\mathcal{H} = \bigoplus_R  \, \mathcal{H}_R$ with $\mathcal{H}_R=\{\vert R,a\rangle, a=1 ...\text{dim }R\}$, where $R$ runs over all unitary irreps of the gauge group $G$. Hence classically, edge degrees of freedom are a set of charges (or currents) that satisfy a Poisson algebra, conjugate to the large gauge transformations that have become physical at the boundary, and that Poisson commute among themselves.

In this work we will find a non-linear generalization of these algebra relations \eqref{eq:1chalg}-\eqref{eq:2chalg}. In particular we will identify a non-linear generalization of the $Q_a$ charge algebra \eqref{eq:1chalg} as describing a Poisson-Lie group with the Poisson-bracket the Semenov-Tian-Shansky bracket. This Poisson algebra is the classical $\hbar \to 0$ limit of the Drinfeld-Jimbo quantum algebra $U_q(\mathfrak{g})$. The conjugate variables $h$ will be shown to satisfy a non-linear ``quadratic'' generalization of \eqref{eq:2chalg}, which will be identified as a Poisson-Lie group with the Poisson bracket the Sklyanin bracket. This Poisson algebra is the classical $\hbar \to 0$ limit of the quantum group $G_q$. The coupling \eqref{eq:3chalg} between both of these Poisson algebras will be deformed as well in terms of what can be called a ``dressing'' bracket. 

\item \textbf{Surface symmetry group $G_s$} \\
When factorizing the state space $\mathcal{H}_{\text{phys}} \hookrightarrow \mathcal{H}_L \otimes \mathcal{H}_R$, a physical state is described by an equivalence class in the extended state space:
\begin{equation}
\ket{v} \otimes \ket{w} \sim \ket{v \cdot g} \otimes \ket{g^{-1} \cdot w}, \qquad \ket{v}\in \mathcal{H}_L, \quad \ket{w} \in \mathcal{H}_R,
\end{equation}
where $g \in G_s$ the surface symmetry group. This relation has geometric meaning in that the right action of $\mathcal{H}_L$ and the left action on $\mathcal{H}_R$ geometrically happen at the cutting (entangling) surface. The fact that a physical state is invariant under this diagonal action, identifies $G_s$ as a gauge symmetry of the full system, with only gauge singlets physical.

We will identify $G_s$ as well as the Poisson-Lie group with the Poisson bracket the Sklyanin bracket. The resulting Poisson algebra is the classical $\hbar \to 0$ limit of the coordinate Hopf algebra of the $q$-deformation of $G$.

\item \textbf{Classification of factorization maps} \\
The edge degrees of freedom need to be complete in the following sense. Given a one-sided theory \eqref{eq:fullone}, including the surface current degrees of freedom, we should be able to glue it to the other side and reproduce the full unsplit theory. In path integral language, this requires the surface charges to be equal as:
\begin{align}
\int \mathcal{D}A_{L\mu} \mathcal{D}J^\mu \mathcal{D}A_{R\mu} \mathcal{D}J'^{\mu} e^{iS_L[A_L] + i\text{Tr}(J^\mu A_{L\mu})}\delta(J^\mu-J'^{\mu })e^{iS_R[A_R] - i\text{Tr}(J'^{\mu} A_{R\mu})}.
\end{align}
In particular, we need a complete set of these surface charges to reobtain the delta functional of \eqref{eq:edgestart} setting $A_L = A_R$ at the boundary $\partial \mathcal{M}$. Various possible ways of consistently factorizing the model exist in general, and it is desired to obtain some notion on how to classify them. These different ways can be unified in terms of ``ungauging'' or ``physicalizing'' would-be-gauge degrees of freedom on the entangling surface.

\end{itemize}

\section{Minimal Factorization of Chern-Simons Theory}

In this section, we describe the structure of the phase space of Chern-Simons theory and factorize it across an entangling surface in a minimal way. We will start with a qualitative argument showing how topological invariance can drastically reduce the number of edge degrees of freedom.

\subsection{Topological Symmetry Reduces Edge Modes}
\label{sec: Topological Symmetry Reduces Edge Modes}

\begin{figure}[h]
    \centering
    \includegraphics[width=0.5\linewidth]{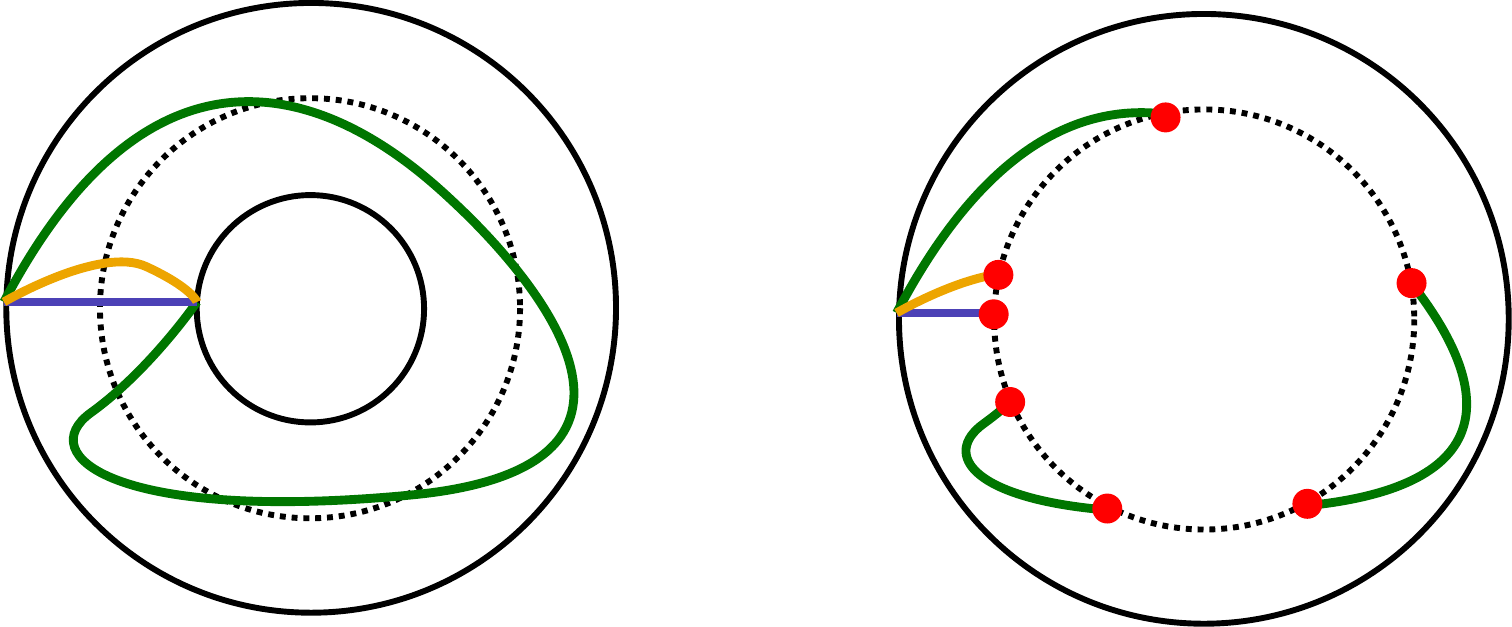}
    \caption{Left: The three colorful lines are Wilson lines anchoring on two physical boundaries (solid black circles). The dashed black circle is the entangling boundary. Right: Outer subsystem with edge modes (red dots) on the entangling boundary.}
    \label{fig: gauge factorization}
\end{figure}

In a 3-dimensional pure gauge theory, degrees of freedom are characterized by Wilson lines. In this work, we focus on a topological gauge theory defined on a 3-manifold with a Cauchy slice of annulus topology, see Fig. \ref{fig: gauge factorization} left. 
The system is divided into two parts (along a circular line) by cutting along an entangling boundary. After cutting, the Wilson line is not invariant under the gauge transformation on the entangling boundary. To restore gauge invariance in the outer subsystem, one can add charges or edge modes on the entangling boundary (see Fig. \ref{fig: gauge factorization} right). Since Wilson lines can intersect any point on the entangling boundary, these edge modes form a field. This is the end of the old story of edge modes.

\begin{figure}[h]
    \centering
    \includegraphics[width=0.5\linewidth]{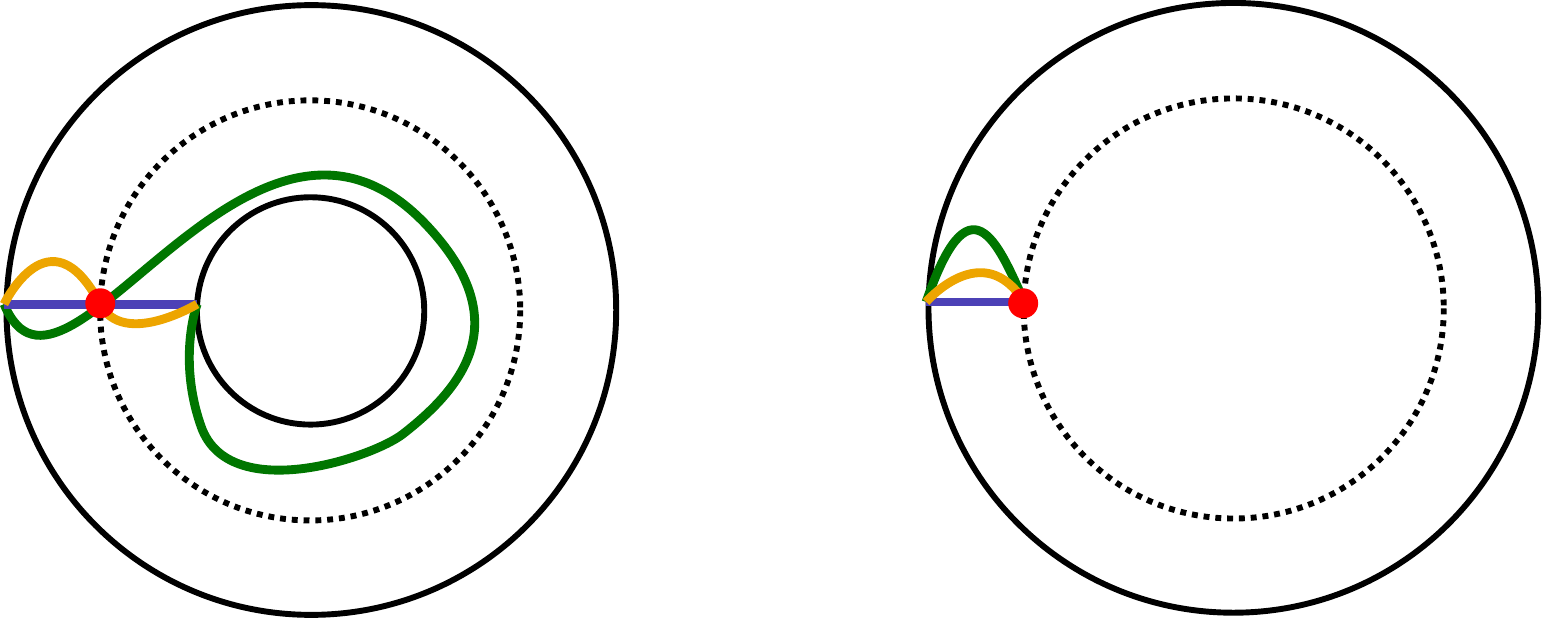}
    \caption{We identify all edge states that only differ by moving the Wilson line endpoint on the entangling surface. All Wilson lines are then equivalently represented to anchor on the same point (red dot) on the entangling boundary.}
    \label{fig: gauge factorization2}
\end{figure}

Now we turn to see how topological invariance can drastically reduce edge degrees of freedom. Within the two-sided theory, we can freely deform all Wilson lines using the topological invariance of the full model. This allows us to force all two-sided Wilson lines to pierce the entangling surface only once and at the same point (see Fig. \ref{fig: gauge factorization2}). From the one-sided perspective, this means all Wilson lines end on the same point without loss of generality. 
What this argument shows is that one does not have to consider other endpoints on the entangling surface, since they are ultimately all equivalent in the two-sided system in a topological gauge theory. Considering them leads to an infinite overcounting of the minimum number of degrees of freedom required to split a topological gauge theory, similar to adding a decoupled qubit to the edge degrees of freedom as we discussed in the Introduction. Such a factorization map reduced by topological invariance was first proposed in lattice gauge theories in \cite{Delcamp:2016eya,Akers:2024wab}, although it was not formulated precisely in the same way. 

\begin{figure}[h]
    \centering
    \includegraphics[width=0.5\linewidth]{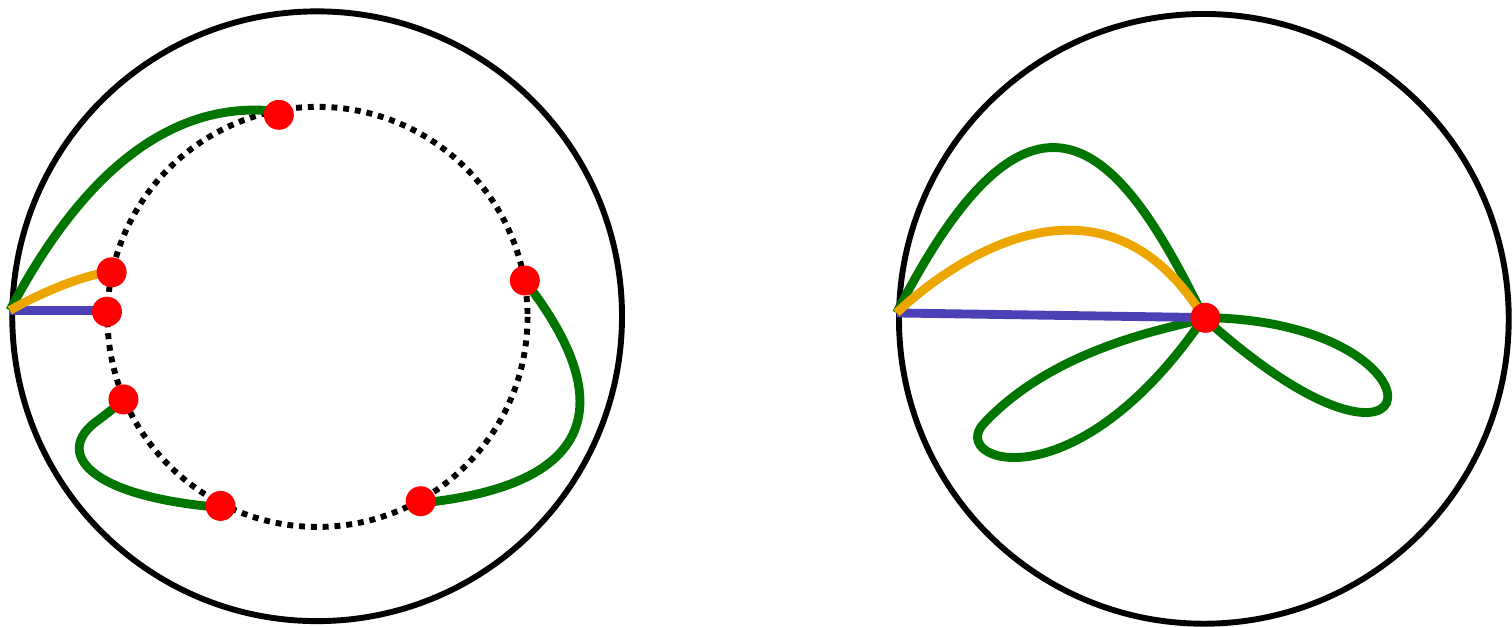}
    \caption{The annulus on the left is homotopy equivalent to a disc with a puncture (red dot). The puncture is a retract of the entangling boundary.}
    \label{fig: gauge factorization3}
\end{figure}

An alternative perspective on how topological invariance reduces edge modes is as follows. In a topological field theory \cite{atiyah1988topological}, specifying the submanifold on which a Wilson line or loop is located is meaningless; only the homology class of the submanifold matters. Since homology is homotopy invariant, we can deform the background manifold without changing physics as long as the homotopy type is kept invariant. In the case of an annulus, this means we can retract the entangling surface to a puncture, as illustrated in Fig. \ref{fig: gauge factorization3}.

\subsection{Phase Space of Chern-Simons Theory on Annulus $\times \,\,\mathbb{R}$}
\label{sec: Phase Space of Chern-Simons Theory}

Consider Chern-Simons theory at level $k$ defined by the Lagrangian
\begin{equation}
\label{eq: bulk chern-simons lagrangian}
    \mathcal{L} = -\frac{k}{4 \pi} \tr (A dA - \frac{2i}{3} A^3),
\end{equation}
defined on Lorentzian signature three-manifolds of the form $S^1 \times I \times \mathbb{R}$ with $S^1$ a circle, $I$ an interval, and $\mathbb{R}$ the real line. This action enjoys the gauge redundancy
\begin{align}
\label{eq: gauge transformation}
     A \to  g A g^{-1} + i g d g^{-1}.
\end{align}
The equation of motion is the Maurer-Cartan equation
\begin{equation}
\label{eq: F=0}
    F = 0,
\end{equation}
with $F$ the field strength
\begin{equation}
    F = dA -i A^2.
\end{equation}
Locally, the equation of motion \eqref{eq: F=0} implies $A$ is pure gauge:
\begin{equation}
\label{eq: local solution}
        A = -i d W W^{-1}
\end{equation}
with $W$ a $G$-valued field that is not necessarily single-valued as one goes around the annulus.

The two boundary components $S^1_R \times \mathbb{R}$ and $S^1_L \times \mathbb{R}$ are cylinders. We denote by $\circledcirc$ the Cauchy slice topology $S^1 \times I$ and choose the orientation
\begin{equation}
    \partial \circledcirc = S_L^1 + S_R^1 .
\end{equation}
We fix coordinates along each boundary to have the same period $2\pi$ for convenience of notation. 
\begin{figure}[h]
    \centering
    \includegraphics[width=0.22\linewidth]{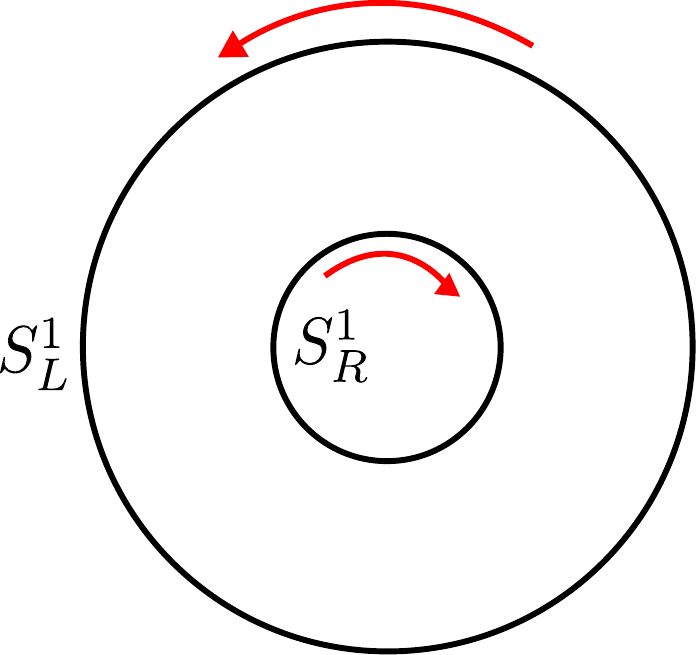}
    \caption{Spatial annulus with orientation of the two boundary circles $S^1_L$ (outer) and $S^1_R$ (inner) depicted.}
    \label{fig:orient}
\end{figure}

There is only one conformal structure on a cylinder. This conformal structure can be characterized by the Hodge star operator $*$ which is conformally invariant and satisfies
\begin{equation}
    ** = 1
\end{equation}
in Lorentzian signature. On the cylinder boundaries, we impose the chiral boundary condition
\begin{align}
    * A|_{S^1_L \times \mathbb{R}} = -A|_{S^1_L \times \mathbb{R}}
    \label{eq: chiral boundary condition +}
    , \qquad 
    * A|_{S^1_R \times \mathbb{R}} = A|_{S^1_R \times \mathbb{R}}.
\end{align} 
reducing the dynamics at each boundary to those of the chiral WZNW model \cite{Elitzur:1989nr}.
Combining the local solution \eqref{eq: local solution} with the chiral boundary conditions \eqref{eq: chiral boundary condition +}, we have
\begin{align}
    A|_{S^1_L \times \mathbb{R}} &= -i \partial W W^{-1},
    \qquad 
    \Bar{\partial} W|_{S^1_L \times \mathbb{R}} = 0,
    \\
    A|_{S^1_R \times \mathbb{R}} &= -i \Bar{\partial} W W^{-1} ,
    \qquad
    \partial W|_{S^1_R \times \mathbb{R}} = 0,
\end{align}
where
\begin{align}
    \partial \equiv \frac{1}{2} ( d - *d ),
    \qquad     \Bar{\partial} \equiv \frac{1}{2} ( d + *d ).
\end{align}

On the equations of motion, the phase space of the model, denoted as $\mathcal{P}_\circledcirc$, can be parameterized by Wilson lines anchoring on the two boundary components.\footnote{Wilson lines with ends on the same boundary can be obtained by combining two Wilson lines on opposite boundaries, and closed Wilson loops vanish.} Using the EOM \eqref{eq: F=0}, a Wilson line $W(x^+,x^-)$ anchoring on $x^+ \in S^1_L$ and $x^- \in S^1_R$ is reduced to a product of operators that only depend on a single boundary coordinate:\footnote{The Wilson line operator $W(x)$, despite depending on a single coordinate, is still non-local since it contains the degree of freedom living at its endpoint at the entangling surface.}
\begin{equation}
    W(x^+,x^-) \equiv\overleftarrow{P} \exp \left( i \int_{x^-}^{x^+} A \right) = W(x^+) W^{-1}(x^-),
\end{equation}
path-ordered as indicated by $\overleftarrow{P}$ with later positions on the left. The flatness condition \eqref{eq: F=0} implies that moving both ends of a Wilson line around the two boundaries once in the same direction respectively has no physical effect, i.e.\footnote{Due to the different orientations chosen of the boundary circles, the LHS has the two factors shift in opposite directions.}
\begin{equation}
    W(x^+ - 2\pi) W^{-1}(x^- + 2\pi) = W(x^+) W^{-1}(x^-).
\end{equation}
Coordinates on the two boundary components are independent, so if we denote the left- and right monodromy as $m_\pm$ respectively:
\begin{align}
    W^{-1}(x^-) W(x^- + 2\pi) = m_-,
    \label{eq: h- def}
    \qquad
   W^{-1}(x^+) W(x^+ + 2\pi) = m_+,
\end{align}
the monodromies are coupled as
    \begin{align}
    m_+ m_- = 1,
    \label{eq: monodromy constraint}
\end{align}
and are independent of $x^\pm$:
\begin{equation}
    d m_\pm = 0.
\end{equation}
Then the phase space $\mathcal{P}_\circledcirc$ as a manifold can be explicitly written as
\begin{equation}
\label{eq: phase sapce annulus}
\boxed{
    \mathcal{P}_\circledcirc = \{ W(x^+) W^{-1}(x^-)| \, W(x^\pm) \in L^{m_\pm} G,\qquad m_\pm \in G, \qquad  m_+ m_- = 1 \}}.
\end{equation}
with $L^m G$ the twisted loop group defined by
\begin{equation}
    L^m G \equiv \{ W:\mathbb{R} \to G |\, W(x+ 2\pi) = W(x) m, \,\,\forall x \in \mathbb{R}  \}.
\end{equation}

This phase space has the following important symmetries:
\begin{itemize}
\item 
Left-multiplication of $W(x^+)$ and right multiplication of $W^{-1}(x^-)$ are just the WZNW Kac-Moody symmetries:\footnote{The right current algebra acts in our notation also from the left, due to our parametrization of $W(x^+) W^{-1}(x^-)$.}
\begin{align}
W(x^+) &\to h(x^+) W(x^+), \qquad  \text{ left current algebra symmetry}, \\
W(x^-) &\to h(x^-) W(x^-), \qquad \text{ right current algebra symmetry}.
\end{align}
\item
There is a gauge redundancy in parametrizing the phase space $\mathcal{P}_\circledcirc $ in terms of the variables $(W(x^+),W(x^-),m_\pm)$, which acts by \emph{right}-multiplication on $W(x^+)$ and \emph{left}-multiplication on $W^{-1}(x^-)$:
\begin{equation}
\label{eq:gauges}
W(x^+) \sim W(x^+)h, \qquad W(x^-) \sim W(x^-)h, \qquad m_\pm \sim h^{-1}m_\pm h,
\end{equation}
for constant group elements $h$.
\end{itemize}

\subsection{Poisson Algebra of Chern-Simons Theory}
\label{sec: Poisson Algebra of Chern-Simons Theory}

The phase space $\mathcal{P}_\circledcirc$ of Chern-Simons theory defined on $\circledcirc \times \mathbb{R}$ with chiral boundary conditions \eqref{eq: chiral boundary condition +} is given by Eq. \eqref{eq: phase sapce annulus}. In this section, we write down the Poisson structure of $\mathcal{P}_\circledcirc$ associated with the Lagrangian \eqref{eq: bulk chern-simons lagrangian}. 

$\mathcal{P}_\circledcirc$ is a space of Wilson lines anchoring on two chiral boundary components,
\begin{equation}
    W(x^+,x^-) = \overleftarrow{P} \exp \left( i \int_{x^-}^{x^+} A \right),
\end{equation}
with $x^\pm \in S^1_{L,R} \times \mathbb{R}$ and with $A$ a flat gauge field. The gauge potential $A$ itself is the conserved current $j$ associated with the gauge symmetry \eqref{eq: gauge transformation} at the chiral boundaries, i.e.
\begin{equation}
\label{eq: j = A = dgg-1}
    j = - \frac{k}{2 \pi} A\vert_{S^1_{L,R} \times \mathbb{R}},
\end{equation}
forming a boundary Kac-Moody current algebra. They have the following action on Wilson lines that hit the boundary:
\begin{align}
    \Big\{  \underset{S^1_L}{\oint} \tr (j \phi ) , W(x^+,x^-) \Big\} &= i \phi(x^+) W(x^+,x^-),
    \label{eq: KM Sym +}
    \\
    \Big\{ \underset{S^1_R}{\oint} \tr (j \phi ) , W(x^+,x^-) \Big\} &= -i  W(x^+,x^-) \phi(x^-),
    \label{eq: KM Sym -}
\end{align}
where $\phi$ is an element in the loop algebra $\{ \phi: S^1 \to \mathfrak{g}|\, \phi(x + 2\pi) = \phi (x), \forall x \in S^1 \}$. Writing the current one-forms on both boundaries as $ j= j(x^+)dx^+$ and $ j= \bar{j}(x^-)dx^-$, the Poisson brackets \eqref{eq: KM Sym +}, \eqref{eq: KM Sym -} can be rewritten as\footnote{
These are analogous to the standard WZNW OPE expansions of the type
\begin{align}
J^a(x_1^+) g(x_2^+) \sim \frac{-T^a g(x^+_1)}{x_1^+-x_2^+}
\end{align}} 
\begin{align}
    \{ j_1({x^+_1}), W_2({x_2^+, y}) \} &= i {C}_{12}{W}_2({x_2^+, y}) \overunderset{\infty}{n=-\infty}{\sum} \delta (x^+_1 - x_2^+ - 2\pi n), \quad y\notin S^1_L,
    \label{eq: KM sym tensor +}
    \\
        \{ \bar{j}_1({x^-_1}), {W}_2({\bar{y}, x^-_2}) \} &= -i {W}_2({ \bar{y}, x^-_2}) {C}_{12} \overunderset{\infty}{n=-\infty}{\sum} \delta (x^-_1 - x^-_2 - 2\pi n ), \quad \bar{y} \notin S^1_R .
    \label{eq: KM sym tensor -}
\end{align}
Here we allow the points $y$ and $\bar{y}$ to be at an arbitrary location, even in the bulk of the system. The ${}_1$ and ${}_2$ subscripts indicate on which tensor factor they act. This is standard notational convention in the classical integrability literature. More precisely, given a basis $\{ T_a\}$ of the Lie algebra $\mg$ of $G$ and an element
    \begin{equation}
        A = A^{ab} T_a \otimes T_b \in \mg^{\otimes 2} ,
    \end{equation}
if $i <j\leq n$, $n \in \mathbb{Z}$, then we define
    \begin{equation}
        {A}_{ij} \equiv A^{ab} \, \mathbf{1}^{\otimes (i-1)} \otimes T_a \otimes \mathbf{1}^{\otimes (j-i-1)} \otimes T_b \otimes \mathbf{1}^{\otimes(n-j)}.
    \end{equation}
For general $A \in \mg^{\otimes m}$, ${A}_{i_1 i_2 \dots i_m}$ is defined similarly. As an example, we have the following equivalent notations, see also \cite{Babelon:2003qtg,grammaticos2004integrability}:\footnote{This is also sometimes denoted in the literature as:
\begin{equation}
\{ j({x^+_1}) \, \overset{\otimes}{,} \, W({x_2^+, x^-}) \} .
\end{equation}
}
\begin{equation}
\{ j_1({x^+_1}), W_2({x_2^+, x^-}) \} \equiv \{ j({x^+_1}) \otimes \mathbf{1}, \mathbf{1} \otimes W({x_2^+, x^-})\}.
\end{equation}
The quantity $C_{12}$ is the tensor quadratic Casimir defined by\footnote{We assume in this work that the Lie algebra $\mathfrak{g}$ is semi-simple such that the inverse $K^{ab}$ exists.}
\begin{align}
    C_{12} \equiv K^{ab}T_a \otimes T_b,
    \qquad
    K_{ab} = \tr (T_a T_b).
\end{align}
This object has the following important properties:
\begin{align}
\label{eq:sym}
&C_{12} = C_{21}, \\
&X_1 = \text{Tr}_2(C_{12}X_2), \\ 
\label{eq:propC}
&\text{Ad}_{h \otimes h} C_{12} \equiv (h \otimes h) C_{12} (h^{-1} \otimes h^{-1})= C_{12}, \quad  \forall X \in \mathfrak{g}, h\in G.
\end{align}
The above algebra \eqref{eq: KM sym tensor +}, \eqref{eq: KM sym tensor -} is equivalent to the Poisson bracket of Wilson lines:
\begin{align}
\label{eq: W, W Poisson KM}
    \{{W}_1(x^+_1,x^-_1) , {W}_2(x^+_2,x^-_2) \} &= \frac{2 \pi}{k} {W}_1(x^+_1,x^-_1) {W}_2(x^+_2,x^-_2) r^{\tx{KM}}_{12} (x^-_{12}, x^+_{12})
    \\
    &=
    -\frac{2\pi}{k} r^{\tx{KM}}_{12} ( x^+_{12}, x^-_{12} ) {W}_1(x^+_1,x^-_1) {W}_2(x^+_2,x^-_2).
\end{align}
See Appendix \ref{app: KM double} for a derivation. We introduced here the object
\begin{equation}
        r^{\tx{KM}}_{12} (x^-_{12}, x^+_{12}) \equiv \overunderset{\infty}{n = - \infty}{\sum}  {W}_1( x^-_1+2\pi n, x^-_2) {C}_{12} {W}_1( x^-_2, x^-_1 +2\pi n) \varepsilon (x^+_{12} + 2n \pi, x^-_{12} + 2n \pi ),
\end{equation}
and we have defined
\begin{equation}
    x^+_{12} \equiv x^+_1 - x^+_2, \quad x^-_{12} \equiv x^-_1 - x^-_2, 
\end{equation}
\begin{equation}
    \tx{sgn}\, x =
\begin{cases}
1, & x \in  (0^+, \infty),
\\
-1, & x \in ( -\infty, 0^-) ,
\end{cases}
\end{equation}
and the oriented intersection number
\begin{equation}
\varepsilon (x^+_{12}, x^-_{12}) \equiv  \frac{1}{2}\left(  \tx{sgn} (x^+_{12}) - \tx{sgn} (x^-_{12})\right),
\end{equation}
which depends on how the Wilson lines intersect as oriented lines. $W(x^-_1 + 2\pi n, x^-_2)$ is a Wilson line starting at $x^-_2$, winding around the annulus $n$ times, and ending at $x^-_1$. 

\begin{figure}[h]
    \centering
\includegraphics[width=0.75\linewidth]{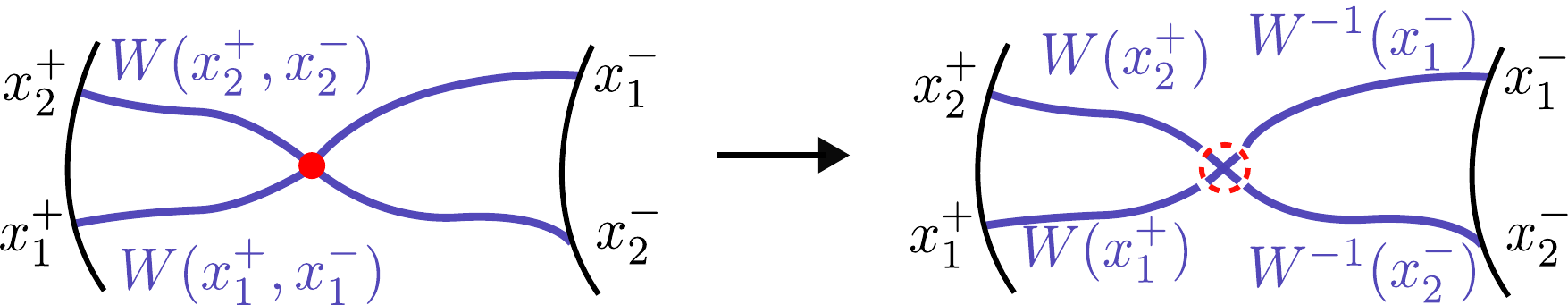}
    \caption{Visualization of the crossing Wilson line algebra with a contribution from the intersection number $\varepsilon (x^+_{12}, x^-_{12})$, by cutting the Wilson lines into pieces.}
    \label{fig:wilsonalg}
\end{figure}

If we restrict to the case of zero winding number, i.e.
\begin{align}
    |x^+_{12}| < 2\pi,
    \quad
    |x^-_{12}| < 2\pi,
\end{align}
then the Poisson bracket \eqref{eq: W, W Poisson KM} reduces to
\begin{align}
\label{eq: w,w poison, KM, n=0}
    &\quad\,\, \{ {W}_1(x^+_1,x^-_1) , {W}_2(x^+_2,x^-_2) \} \nonumber \\ 
    &= \frac{\pi}{k} {W}_1(x^+_1,x^-_1) {W}_2(x^+_2,x^-_2) ({W}_1(x^-_1, x^-_2) {C}_{12} {W}_1(x^-_2, x^-_1))[\tx{sgn}(x^+_{12}) - \tx{sgn}(x^-_{12} ) ]
    \\
    &=
     \frac{\pi}{k} ({W}_2(x^+_2, x^+_1) {C}_{12} {W}_2(x^+_1, x^+_2) ){W}_1(x^+_1,x^-_1) {W}_2(x^+_2,x^-_2) [\tx{sgn}(x^+_{12}) - \tx{sgn}(x^-_{12} )] , 
\end{align}
where the equality of the two lines implements the swapping symmetry between both boundaries. It can be very explicitly checked by writing $W(x^+,x^-)=W(x^+) W^{-1}(x^-)$ and using \eqref{eq:propC}, to write either of the two expressions directly as
\begin{equation}
\label{eq:alglast}
\{{W}_1(x^+_1,x^-_1) , {W}_2(x^+_2,x^-_2) \} = \frac{2\pi}{k} W_1(x_1^+)W_2(x_2^+) C_{12} W_1^{-1}(x_1^-)W_2^{-1}(x_2^-)\varepsilon (x^+_{12}, x^-_{12}).
\end{equation}
This Wilson line Poisson algebra is known in the literature, see e.g. section 4.7 in the textbook \cite{Carlip:1998uc}, where in particular our last way of writing the algebra \eqref{eq:alglast} is directly matched by cutting up two crossing Wilson lines as illustrated in Fig.~\ref{fig:wilsonalg}.\footnote{The infinitesimal segments in the middle can be Taylor-expanded 
\begin{equation}
\overleftarrow{P}\exp\left(i\int_{C_\epsilon} A\right) \approx \mathbf{1} + i\int_{C_\epsilon} ds\,A_i^a T_a \frac{dx^i}{ds},
\end{equation}
and one finds the same answer \eqref{eq:alglast} upon using the standard gauge field canonical commutator (when identifying the degrees of freedom as the spatial components $A_i$):
\begin{equation}
\{A_i^a(\vec{x}),A_j^b(\vec{y})\} = -\frac{2\pi}{k}\epsilon_{ij}K^{ab}\delta(\vec{x}-\vec{y}), \qquad K_{ab}=\text{Tr}(T_aT_b),
\end{equation}
and recognizing the geometric intersection number in the remaining integrals.}
The above algebra \eqref{eq:alglast} can be further simplified on a case-by-case basis using the skein relations of the particular group $G$, as is usually done for the particular case of SL$(2,\mathbb{R})$.\footnote{This type of Poisson algebra for closed Wilson loops was studied by Goldman \cite{Goldman1986InvariantFO} and quantized by Turaev \cite{Turaev1991}.}

\subsection{Phase Space of $\sqrt{\tx{Chern-Simons}}$ Theory}
\label{s:sqrch}
We now consider Chern-Simons theory on the one-sided geometry bounded by one physical WZNW boundary and one entanglement boundary. As explained in Section \ref{sec: Topological Symmetry Reduces Edge Modes}, the physics is only sensitive to the homotopy type of the entangling boundary, so we can simply set the topology of the Cauchy slice to be a disc with a puncture in the bulk.\footnote{Technically, the puncture comes equiped with a cilium. This means one can distinguish between different Wilson lines attached to the point in terms of their winding number around the puncture. This is e.g. illustrated later in Fig.~\ref{fig:monodromy}.} One can also directly see this equivalence from Eq. \eqref{eq: phase sapce annulus}, i.e. the full phase space $\mathcal{P}_\circledcirc$ does not depend on the size of the entangling boundary. As will become clear in Section \ref{sec: poisson algebra of sqrt CS}, we call this theory the $\sqrt{\tx{Chern-Simons}}$ theory.

\begin{figure}[h]
    \centering
\includegraphics[width=0.3\linewidth]{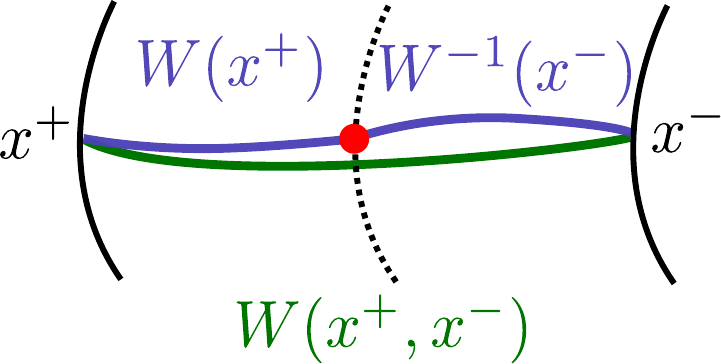}
    \caption{Splitting a Wilson line $W(x^+,x^-)$ into two one-sided non-local Wilson lines $W(x^+)$ and $W^{-1}(x^-)$ that end on the entangling surface (dashed line).}
    \label{fig:wilsonsplit}
\end{figure}

We denote by $\odot$ a disc with a puncture in the bulk, by $W(x)$ the Wilson line from the puncture to a point $x$ on the boundary cylinder, and by $m$ the monodromy\footnote{Since we focus solely on a single boundary, we drop the $\pm$ superscripts in this and the next subsection to avoid cluttering the equations.}
\begin{equation}
\label{eq: monodromy}
    m \equiv W^{-1}(x-2\pi) W(x).
\end{equation}
Then the phase space of the Chern-Simons theory on the punctured disc $\odot$ is\footnote{\label{ft: topo B.C.}A priori, the monodromy $m$ could be dynamical. Since the puncture is physically the entangling surface, and the time evolution is done using the one-sided Hamiltonian, it is natural to impose the following boundary condition on the puncture to remove dynamics inaccessible to the one-sided observer:
\begin{equation}
\label{eq: topological B.C.}
    \partial_u W(x,u) = 0.
\end{equation}
An alternative motivation to introduce Eq. \eqref{eq: topological B.C.} is that the topological invariance implies the bulk Hamiltonian is zero. If we don't impose any boundary condition on the puncture, the corresponding phase space will not give rise to a minimal extension due to the redundant dynamics on the puncture.}
\begin{equation}
    \mathcal{P}_{\odot} = \underset{m \in G }{\bigcup} L^{m}G.
\end{equation}
We can recover a Wilson line anchored on the two outer boundaries by gluing the Wilson lines on the two punctured discs, i.e.
\begin{equation}
\label{eq:split}
    W(x^+, x^-) = W(x^+) W^{-1}(x^-),
\end{equation}
visualized in Fig. \ref{fig:wilsonsplit}.

This leads to the gluing map:\footnote{Geometrically, an annulus can be obtained by gluing together the punctures of two punctured discs. More precisely, the surface obtained in this way is not homeomorphic but homotopy equivalent to an annulus. So this factorization is impossible for theories without topological invariance as discussed in Section \ref{sec: Topological Symmetry Reduces Edge Modes}.}
\begin{alignat}
{4}
\; && \mathcal{P}_{ \odot }  \times \mathcal{P}_{ \odot} && \;\twoheadrightarrow\; & \mathcal{P}_\circledcirc,
\label{eq: gluing morphism}\\
          && \left(W(x^+), W(x^-)\right)\vert_{m_+ m_- = \mathbf{1}}  \,\,  &&  \;\mapsto\; & \,\, W(x^+) W^{-1}(x^-) \vert_{m_+ m_- = \mathbf{1}}.
\end{alignat} 
The phase space $\mathcal{P}_\circledcirc$ can then be identified as
\begin{align}
\label{eq: coimage isomorphism}
        \mathcal{P}_\circledcirc 
        \cong \frac{1}{G} \underset{m \in G}{\bigcup} L^mG \times L^{m^{-1}}G ,
\end{align}
with the quotient induced by the right $G$-action on $\mathcal{P}_{ \odot }  \times \mathcal{P}_\odot$,
\begin{alignat}{4}
\; && G \times \mathcal{P}_\odot  \times \mathcal{P}_\odot && \;\to\;     & \mathcal{P}_\odot  \times \mathcal{P}_\odot,
\label{eq: Poisson-Lie action}
\\
     && (h, W(x^+), W(x^-))    && \;\mapsto\; & (W(x^+) \cdot h, W(x^-)\cdot  h).
\end{alignat}
The right $G$-action \eqref{eq: Poisson-Lie action} is actually a Poisson-Lie symmetry~\cite{Semenov-Tian-Shansky:1985mgd}, as we will explain later in Section \ref{s:prop2}. Since the right $G$-action \eqref{eq: Poisson-Lie action} is modded out in Eq. \eqref{eq: coimage isomorphism} for the two-sided space, this Poisson-Lie symmetry is a gauge symmetry in $\mathcal{P}_\circledcirc$, and a physical observable is in the singlet sector of this symmetry. 

\subsection{Poisson Algebra of $\sqrt{\tx{Chern-Simons}}$ Theory}
\label{sec: poisson algebra of sqrt CS}
In this section, we show that the moduli space of minimally extended phase spaces of the Chern-Simons theory is the solution space of the modified classical Yang-Baxter equation. In Section \ref{sec: Poisson Algebra of Chern-Simons Theory}, we showed that the Poisson algebra on $\mathcal{P}_\circledcirc$ is \eqref{eq: KM sym tensor +}, \eqref{eq: KM sym tensor -} associated with the two physical boundaries. For $\mathcal{P}_\odot$, we only have one physical boundary, and hence only a single copy of this algebra:
\begin{equation}
\label{eq: KM sym sqrt}
                \{{j}_1({x_1}), {W}_2(x_2) \} = i {C}_{12} {W}_2(x_2) \overunderset{\infty}{n=-\infty}{\sum} \delta (x_{12} - 2\pi n), 
\end{equation}
with $x_{12} = x_1 - x_2$, and where we used \eqref{eq:split}. Unlike the case of the algebra \eqref{eq: KM sym tensor +}, \eqref{eq: KM sym tensor -} which led to Eq. \eqref{eq: W, W Poisson KM}, the single-sided case Eq. \eqref{eq: KM sym sqrt} does \emph{not} fully fix the Poisson structure of the $W$'s on $\mathcal{P}_\odot$ (as we will show below) since the degrees of freedom on the puncture/entangling surface are irrelevant for the Kac-Moody symmetry on the physical (chiral) boundary. Thus we need to complete the single-sided algebra by introducing additional data in the Poisson bracket associated with the puncture.
This mathematical observation corresponds physically to the need to making a choice of edge state description of the single-sided system.

The most general way to complete the Poisson algebra on $\mathcal{P}_\odot$ is to note that Eq. \eqref{eq: KM sym sqrt} can be rewritten as a differential equation of $\{{W}_1(x), {W}_2(y) \}$, i.e.\footnote{See also \cite{balog2000chiral} for a related method. Another approach is to consider the lattice regularization of the Kac-Moody algebra~\cite{Alekseev:1991wq, Alekseev:1992wn}.} \footnote{A useful identity to find this formula is 
\begin{equation}
\{g^{-1},F\} = - g^{-1}\{g,F\}g^{-1},
\end{equation}
which holds in any representation. One also needs to use the properties of the tensor Casimir \eqref{eq:propC}.
}
\begin{align}
\label{eq: d-eq for W,W}
        \partial_{x_1} \Big({W}^{-1}_1({x_1}) \{ {W}_1({x_1}), {W}_2({x_2}) \} \Big) 
    =
    \frac{\pi}{k} \overunderset{\infty}{n=- \infty}{\sum} {m}_1^{-n} { W}_2({x_2}) {C}_{12} {m}_1^n \partial_{x_1} \tx{sgn} (x_{12} - 2\pi n).
\end{align}
Integrating both sides of Eq.\eqref{eq: d-eq for W,W} and using the antisymmetry of the Poisson bracket, we get
\begin{equation}
\label{eq: W,W poisson, winding}
        \{{W}_1(x_1), {W}_2(x_2) \} = \frac{2 \pi}{k} {W}_1(x_1) {W}_2(x_2) \Big({\tilde{r}}_{12} +  \frac{1}{2} \overunderset{\infty}{n=- \infty}{\sum}  \tx{sgn}(x_{12} - 2\pi n ) {m}_1^{- n} {C}_{12} {m}_1^{ n} \Big) ,
\end{equation}
with $\tilde{r} \in \mg \otimes \mg $ the integration constant. The antisymmetry of the Poisson bracket implies\footnote{The last term on the RHS of \eqref{eq: W,W poisson, winding} is antisymmetric as can be seen using \eqref{eq:sym}, \eqref{eq:propC}, and changing $n \to -n$.}
\begin{equation}
{\tilde{r}}_{12} =- \tilde{r}_{21}.
\end{equation}
If $ |x_{12}| < 2\pi$ (i.e. zero winding), then the Poisson bracket \eqref{eq: W,W poisson, winding} reduces to
\begin{equation}
\label{eq: W, W poisson, r}
\boxed{
    \{ {W}_1(x_1), {W}_2(x_2) \} = \frac{2 \pi}{k} {W}_1(x_1) {W}_2(x_2) {r}_{12}(x_{12})},
\end{equation}
where
\begin{align}
     {r}_{12}(x_{12}) \equiv  {r}_{12} + \frac{1}{2}  \,  {C}_{12}\, \tx{sgn}(x_{12}),
    \qquad
    {r}_{12} \equiv {\tilde{r}}_{12} - \frac{1}{2} \underset{n \ne 0}{\sum}  {m}_2^{n}   {C}_{12}\, {m}_2^{ -n}  \tx{sgn}( n ),
    \label{eq: classical r-matrix}
\end{align}
and we have bundled together in $r_{12}$ all $x_{12}$-independent numbers. We can just as well view $r_{12}$ as the integration constant, which is also antisymmetric: $r_{12}=-r_{21}$.

A Poisson bracket of the form of Eq. \eqref{eq: W, W poisson, r} was discovered in the WZNW model by Faddeev \cite{faddeev1990exchange}. Poisson brackets of similar form were discovered earlier in other integrable models, see e.g. \cite{faddeev1982integrable}. Brackets of this type have been (somewhat ad hoc) proposed in \cite{Fock:1998nu,Alekseev:1994pa,Alekseev:1994au,Han:2025xfl} to describe the combinatorial quantization of Chern-Simons theory. We have instead bootstrapped their form directly by integrating \eqref{eq: KM sym sqrt}, crucially relying on the additional presence of a holographic (WZNW) boundary.

After quantization, the Poisson bracket \eqref{eq: W, W poisson, r} will yield a braiding algebra, as we briefly discuss later in section \ref{s:quant}.

Since we are seeking a \emph{minimal} extension of the model that allows a proper factorization of the phase space across the entangling surface, we will assume that the integration constant $r_{12}$ is independent of the monodromy $m$.\footnote{Without this assumption, we would be led to the notion of Poisson-Lie groupoid \cite{Balog:1999wy} which is much more complicated and beyond the scope of this paper. It would presumably lead to a consistent but more involved factorization map.} Crucially, the Poisson bracket \eqref{eq: W, W poisson, r} satisfies the Jacobi identity if and only if $r_{12}$ in Eq. \eqref{eq: classical r-matrix} satisfies the modified classical Yang-Baxter equation (\textbf{MCYBE}) with negative coefficient on the RHS:\footnote{This is also called the modified classical Yang-Baxter equation of split type.}
\begin{equation}
\label{eq: MCYBE}
    [{r}_{12}, {r}_{23}] + [{r}_{23}, {r}_{31}] + [{r}_{31}, {r}_{12}] = -\frac{1}{4} f,
\end{equation}
where
\begin{align}
    f \equiv f^{abc} T_a \otimes T_b \otimes T_c,
\qquad
    f^{abc} = K^{ad} K^{be} f_{de}^{\,\,\,\,\, c},
    \qquad
    [T_a, T_b] = f_{ab}^{\,\,\,\,\, c}\, T_c.
\end{align}

A solution of the MCYBE \eqref{eq: MCYBE} is called a classical $r$-matrix. Since the minimally extended phase space of the full two-sided system is $\mathcal{P}_\odot \times \mathcal{P}_\odot$, one may naively conclude an extension is labelled by two classical $r$-matrices, one on each side of the entangling surface. However, the monodromy coupling \eqref{eq: monodromy constraint} of the two sides makes the two classical $r$-matrices also 1:1 related. We can see this directly from the Poisson brackets involving the monodromy as follows. Suppose $-2 \pi < y < x < 2 \pi$, then we have:
\begin{align}
    \{ {m}_1, {W}_2(x) \}
    &=
    \{ {W}^{-1}_1(y) W_1(y+2\pi), {W}_2(x) \}
    \\
    &=
    \{ {W}^{-1}_1(y), {W}_2(x) \} W_1(y+2\pi) + {W}^{-1}_1(y) \{ W_1(y+2\pi), {W}_2(x) \}
    \\
    &=
    -{W}^{-1}_1(y)\{ {W }_1(y), {W}_2(x) \} {m}_1 + {W}^{-1}_1(y) \{W_1(y+2\pi), {W}_2(x) \}.
\end{align}
Using Eq. \eqref{eq: W, W poisson, r}, we reach
\begin{align}
\label{eq: dressing action}
        \boxed{\{ m_1, {W}_2(x) \}
    = \frac{2 \pi}{k} {W}_2({x}) ( {m}_1 r^+_{12} - r^-_{12} {m}_1 )},
\end{align}
where we introduced the standard notation
\begin{equation}
    r^\pm \equiv r \pm \frac{1}{2} C,
\end{equation}
and used $W_1(y+2\pi) = W_1(y)m_1$. As is well-known \cite{Babelon:2003qtg}, if $r$ satisfies the MCYBE \eqref{eq: MCYBE}, the $r^\pm$-matrices satisfy the classical Yang-Baxter equation:
\begin{equation}
\label{eq:CYBE}
    [{r}^\pm_{12}, {r}^\pm_{23}] + [{r}^\pm_{23}, {r}^\pm_{31}] + [{r}^\pm_{31}, {r}^\pm_{12}] = 0.
\end{equation}
The Poisson bracket \eqref{eq: dressing action} describes how the monodromy acts on Wilson lines. Using the Leibniz rule, we can finally derive the Poisson bracket of monodromies:
\begin{align}
    \big\{{m}_1, {m}_2  \big\}
    &=
    \big\{ {m}_1,{ W^{-1}_2(x) W_2(x+2\pi) } \big\} \nonumber
    \\
    &=
    { W^{-1}_2(x)} \big\{ {m}_1, W_2(x+2\pi) \big\} + \big\{ {m}_1, { W}_2^{-1}(x)   \big\} W_2(x+2\pi) \nonumber
    \\
    &=
    W_2^{-1}(x) \big\{ {m}_1, W_2(x+2\pi)\big\} - W_2^{-1}(x)  \big\{{m}_1, { W_2(x)  } \big\} W_2^{-1}(x)  W_2(x+2\pi) \nonumber
    \\
    &=
    \frac{2 \pi}{k} W_2^{-1}(x)  W_2(x+2\pi) ( {m}_1 r^+_{12} - r^-_{12} {m}_1 ) - \frac{2 \pi}{k} ( {m}_1 r^+_{12} - r^-_{12} {m}_1 ) W_2^{-1}(x)   W_2(x+2\pi) \nonumber
    \\
    &=
            \frac{2 \pi}{k} (r^-_{12}{m}_1 {m}_2 - {m}_1 r^+_{12} {m}_2 - {m}_2 r^-_{12} {m}_1 +  {m}_1 {m}_2 r^+_{12}).
\end{align}
Using the identity\footnote{This can be readily proven using eq. \eqref{eq:propC}.}
\begin{equation}
{m}_1 {m}_2 r^+_{12} + r^-_{12} {m}_1 {m}_2 =  r^+_{12}{m}_1 {m}_2 + {m}_1 {m}_2 r^-_{12},
\end{equation}
we finally have
\begin{equation}
\label{eq: STS bracket}
\boxed{
     \{{m}_1, {m}_2  \} = \frac{2 \pi}{k} (r^+_{12} {m}_1 {m}_2 - {m}_1 r^+_{12} {m}_2 - {m}_2 r^-_{12} {m}_1 +  {m}_1 {m}_2 r^-_{12})}.
\end{equation}
Eq. \eqref{eq: STS bracket} is called the Semenov-Tian-Shansky (\textbf{STS}) bracket \cite{Semenov-Tian-Shansky:1985mgd, semenov1992poisson}. Note that the trace of the monodromy matrix $\text{Tr}(m)$ is a Casimir function of this Poisson algebra since $\{\text{Tr}(m_1),m_2\}=0$ as readily checked.

According to the monodromy coupling \eqref{eq: monodromy constraint}, the monodromy on a chiral factor $\mathcal{P}_\circledcirc$ is $m^{-1}$ if the other is $m$. The corresponding Poisson bracket of the other sector is hence
\begin{equation}
    \{{m}_1^{-1}, {m}_2^{-1}  \} = \frac{2 \pi}{k} ( 
r^-_{12}{m}_1^{-1} {m}_2^{-1} - {m}_1^{-1} r^-_{12} {m}_2^{-1} - {m}_2^{-1}  r^+_{12} {m}_1^{-1}  +  {m}_1^{-1} {m}_2^{-1} 
 r^+_{12} ),
\end{equation}
which identifies the $r$-matrix on the other side simply as the Cayley transform $r^\pm \leftrightarrow r^\mp$.

Since the classical $r$-matrices of the two chiral factors $\mathcal{P}_\odot$ are coupled, \emph{the moduli space of minimally extended phase spaces is precisely the solution space of the MCYBE} \eqref{eq: MCYBE}.\footnote{As a philosophical comment, by taking the square root of the Klein-Gordon equation, Dirac did not find any specific $\gamma$-matrices, but discovered the Clifford algebra. Similarly, by taking the square root of the Poisson algebra of Chern-Simons theory, we don't find a specific $r$-matrix, but discover the modified classical Yang-Baxter equation.}

There is an alternative route towards precisely this same structure and statement, focusing instead on the symplectic form, the inverse of the Poisson algebra we discussed above. In that case we can follow the early literature on factorization of the non-chiral WZNW model into its chiral sectors to provide a complementary perspective on the above results \cite{Gawedzki:1990jc,Alekseev:1990vr,Falceto:1992bf}. We collect and review how this argument works in Appendix \ref{app:symlform}.

\section{Nonlinear Edge Charge Algebra}
\label{s:prop1}
In this section, we explicitly construct the (nonlinear) algebra satisfied by the classical edge degrees of freedom.

\subsection{Poisson Algebra of a Particle on a Quantum Group}
\label{s:poqg}
In Section \ref{sec: poisson algebra of sqrt CS}, we have obtained and classified the minimal factorization map \eqref{eq: gluing morphism} describing how the one-sided systems (the $\sqrt{\tx{Chern-Simons}}$ theories) are combined into the two-sided system as $\mathcal{P}_{ \odot }  \times \mathcal{P}_{ \odot}  \;\twoheadrightarrow\;  \mathcal{P}_\circledcirc$. In this section, we show that the Poisson algebra \eqref{eq: W,W poisson, winding} of the $\sqrt{\tx{Chern-Simons}}$ theory decomposes as the Poisson algebra of a chiral WZNW model coupled with a particle on a quantum group, as schematically summarized in Eq. \eqref{eq: sqrt CS = chi WZNW times PQG}.

\begin{figure}[h]
    \centering
\includegraphics[width=0.2\linewidth]{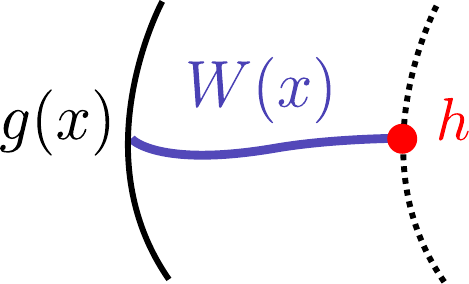}
    \caption{We decompose the non-local operator $W(x)$ into two local group-valued contributions $g(x)$ and $h$ localized on an outer boundary, and the entangling surface respectively.}
    \label{fig:wilsonsplit2}
\end{figure}

In $\mathcal{P}_\odot$, the Wilson line $W(x)$ can be written as the product of two local operators
\begin{equation}
\label{eq: W=gb}
    W(x) = g(x) h
\end{equation}
with $g(x)$ and $h$ located on the physical boundary and the entangling surface (or puncture) respectively (see Fig. \ref{fig:wilsonsplit2}).\footnote{Note that $g(x)$ is not single-valued, but it is localized on the boundary.} 

\begin{figure}[h]
    \centering
\includegraphics[width=0.45\linewidth]{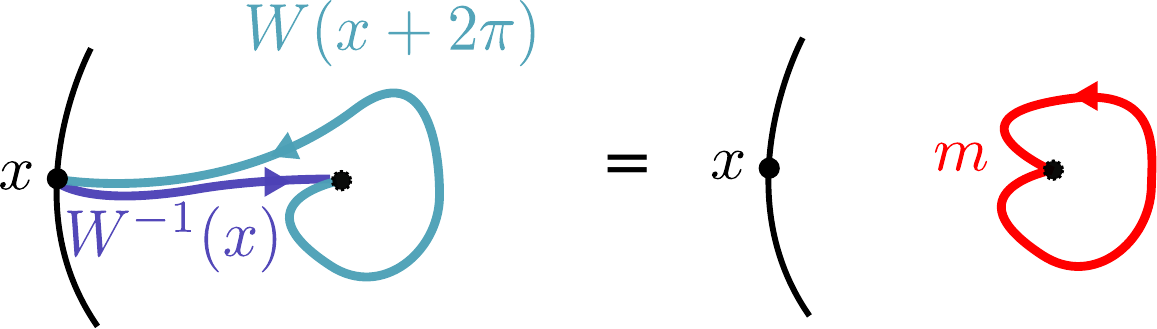}
    \caption{The monodromy variable $m$ is not spatially connected to the outer boundary, but is a winding one loop around the entangling surface (here represented shrunk down to a puncture). Note that we need to keep track of a marked point on the puncture to distinguish physically different monodromies $m$.}
    \label{fig:monodromy}
\end{figure}

The degree of freedom $h$ can be interpreted as the gauge Poisson-Lie symmetry \eqref{eq: Poisson-Lie action} physicalized on the entangling surface. Since $g(x)$ and $h$ live on distinct boundaries, we impose locality
\begin{equation}
\label{eq: locality}
    \{ g(x), h \} = 0.
\end{equation}
Moreover, $\{g_{1}(x_1), g_{2}(x_2)\}$ satisfies the same differential equation \eqref{eq: d-eq for W,W}. The same equation has the same solution, so
\begin{equation}
\label{eq: g, g poison r}
    \{ {g}_{1}(x_1), {g}_{2}(x_2) \} = \frac{2 \pi}{k} {g}_{1}(x_1) {g}_{2}(x_2) {r}_{12}(x_{12}), \quad |x_{12}|< 2\pi .
\end{equation}
with naturally the same classical $r$-matrix by continuity (or gauge redundancy).\footnote{We make some more comments on this in Appendix \ref{app:comments}.} Eq. \eqref{eq: g, g poison r} produces the Poisson algebra of a chiral WZNW model, identified a long time ago \cite{faddeev1990exchange}. Combining Eq. \eqref{eq: W, W poisson, r}, \eqref{eq: W=gb}, \eqref{eq: locality} and \eqref{eq: g, g poison r}, we obtain
\begin{equation}
\label{eq: b,b poisson}
\boxed{
    \{ h \otimes 1, 1 \otimes h \} = \frac{2 \pi}{k} [h \otimes h, r]}.
\end{equation}
The Poisson bracket \eqref{eq: b,b poisson} is called the Sklyanin bracket \cite{sklyanin1979complete, sklyanin1982some, sklyanin1983some, faddeev1982integrable}. A Lie group equipped with a Sklyanin bracket is a Poisson-Lie group.\footnote{A compatible Poisson structure on a Lie group $G$ is determined by a cocycle in the first cohomology group of the Lie algebra of $G$. Since the first Lie algebra cohomology of any semisimple Lie algebra is trivial, a compatible Poisson structure is determined by a 1-coboundary. The Poisson structure determined by a 1-coboundary is called the Sklyanin bracket and thus is the most general Poisson structure one can define on a semisimple Lie group.} \footnote{A Poisson-Lie group is the semiclassical limit of a quantum group. The commutator $[,]$ on a quantum group is related to the Poisson bracket $\{,\}$ on the corresponding Poisson-Lie group via
\begin{equation}
\label{eq: classical limit}
    \underset{\hbar \rightarrow 0}{\tx{lim}}   \frac{1}{i\hbar} [,]  = \{,\} \circ \underset{\hbar \rightarrow 0}{\tx{lim}} 
\end{equation}
with $\hbar$ Planck's constant.} The Sklyanin bracket \eqref{eq: b,b poisson} is the Poisson bracket of the coordinates on a Poisson-Lie group. The corresponding momentum is the monodromy $m$ in Eq.\eqref{eq: monodromy}. The reason is as follows. First, the monodromy is indeed a local observable on the puncture since the field strength is zero according to the equation of motion \eqref{eq: F=0} (and we can hence contract it to the puncture). In pictures (Fig.~\ref{fig:monodromy}), we can deform the Wilson lines defining $m$ to prove that it commutes with local fields on the outer boundaries: $\{m,g(x)\}=0$.

Second, the monodromy generates a "translation" of the coordinate $h$. More precisely, combining Eq. \eqref{eq: dressing action}, \eqref{eq: W=gb}, and \eqref{eq: locality}, we have
\begin{equation}
\label{eq: m, b dressing transform}
\boxed{
    \{ m_1, h_2 \}
    = \frac{2 \pi}{k}  h_2\, ({m}_1 r^+_{12} - r^-_{12} {m}_1 )}.
\end{equation}
As we will explain below, this "translation" reduces to the right $G$-multiplication in the $k \to \infty$ limit. 

The brackets \eqref{eq: STS bracket}, \eqref{eq: b,b poisson}, and \eqref{eq: m, b dressing transform} combined define a consistent Poisson structure (i.e. non-trivially satisfying the Jacobi identity) on the Drinfel'd double $DG$ of $G$~\cite{Drinfeld:1986in}, called usually the \emph{Heisenberg double}.\footnote{The Drinfel'd double as a manifold admits different Poisson structures. The Poisson structure we derived here is called the Heisenberg double \cite{Alekseev:1993qs} or symplectic double \cite{semenov2008integrable}. The Drinfel'd double is also sometimes called the classical double \cite{Babelon:2003qtg} or just double \cite{kosmann2007lie} in the literature.} 
We can rephrase this as 
\emph{the phase space of a particle on a quantum group is the Heisenberg double of the Poisson-Lie group equipped with the Sklyanin bracket \eqref{eq: b,b poisson}}, agreeing with similar statements in \cite{Alekseev:1993qs}. 

The Heisenberg double reduces to the cotangent bundle $T^*G$ of $G$ in the undeformed limit $k\to \infty$ \cite{Alekseev:1993qs}, which is the phase space of a particle on an ordinary group. In detail, defining the current $J$ in terms of the monodromy $m$
\begin{equation}
m = e^{-\frac{2\pi}{k}J},
\end{equation}
in the limit $k \to \infty$, the brackets eq. \eqref{eq: STS bracket}, \eqref{eq: m, b dressing transform} and \eqref{eq: b,b poisson} will reduce to the Poisson algebra of a particle on an ordinary Lie group,
\begin{align}
    \{ {J}_1 , {J}_2 \} &= [{J}_1, {C}_{12} ],
    \label{eq: J, J, poison, r=0}
    \\
    \{{J}_1, {h}_2\} &= - {h}_2\,{C}_{12},
    \label{eq: j, b , poison, r=0}
    \\
    \{ {h}_1, {h}_2\} &= 0
    \label{eq: b, b= 0}.
\end{align}
This Poisson algebra \eqref{eq: J, J, poison, r=0}, \eqref{eq: j, b , poison, r=0} and \eqref{eq: b, b= 0} can be written in component form as
\begin{align}
    \{ J_a, J_b \} &=  f_{ab}^{\,\,\,\,\,c} J_c,
    \\
    \{ J_a, h_{ij} \} &= - (h T_a)_{ij},
    \\
    \{ h_{i j}, h_{i' j'}\} &= 0.
\end{align}
where the $J_a \equiv \tr (T_a J)$ is the charge component in terms of the basis $\{T_a\}$ of $\mg$. The 
$h_{ij}$'s are matrix entries of $h$ in any representation.

Similarly, if we take the $k \to \infty$ limit of the Poisson algebra of Wilson lines \eqref{eq: W, W Poisson KM} of Chern-Simons theory, we recover the Poisson algebra of 2d Yang-Mills theory. The relation between 2d Yang-Mills theory and Chern-Simons theory in the $k \to \infty$ limit was first observed in \cite{Witten:1991we}.

\subsection{Application: U(1) Edge Modes}
As a first (almost trivial) example of the edge algebra we constructed above in \eqref{eq: STS bracket}, \eqref{eq: b,b poisson}, \eqref{eq: m, b dressing transform}, we write it down for the U(1) case. Since the Lie algebra is abelian, a solution $r$ to the MCYBE \eqref{eq: MCYBE} is a complex number. Since both $m$ and $h$ are now $1\times 1$ matrices, we obtain the algebra:
\begin{align}
     \{{m}, {m}  \} &= 0, \\
     \{ h, h \} &= 0, \\
     \{ m, h \}
    &= \frac{2 \pi}{k}  h{m},
\end{align}
where we dropped the superfluous tensor indices, and the $r$-matrix is irrelevant. Identifying 
$m=e^{i\frac{2\pi}{k}p}$ and $h=e^{iq}$, 
the last relation becomes
\begin{equation}
\{q,p\} = 1,
\end{equation}
directly identifying the edge algebra as containing a single degree of freedom $q$, an abelian charge at the entangling surface, and its conjugate $p$. Note that whether $k$ is finite or taken in the linear limit ($k\to +\infty)$, the structure of this edge algebra is the same.

\subsection{Application: Gravitational Anyonic Edge Modes}
\label{s:sl2}

To make the above construction more explicit in a non-trivial case, we evaluate all Poisson brackets for the Heisenberg double of SL$(2,\mathbb{R})$. Our motivation is both as an illustration, and with $\Lambda <0$ Lorentzian 3d gravity in mind since it can be written in its first-order formulation as SL$(2,\mathbb{R})$ $\times$ SL$(2,\mathbb{R})$ Chern-Simons theory \cite{Witten:1988hc,Achucarro:1986uwr}. We will only describe a single copy of SL$(2,\mathbb{R})$ here, meaning that one has to double the resulting algebra if one actually wants to describe 3d gravity. We go through all three Poisson bracket relations. \\

\textbf{Sklyanin bracket \eqref{eq: b,b poisson}:} \\
With the explicit form of the quasi-triangular $r$-matrix:
\begin{equation}
r = \frac{1}{2}\left[\begin{array}{cccc}
0 & 0 & 0 & 0 \\
0 & 0 & -1 & 0 \\
0 & 1  & 0 & 0 \\
0 & 0 & 0 & 0
\end{array}\right], \qquad
r^\pm \equiv r \pm \frac{1}{2}C_{12} = \pm\frac{1}{4}\left[\begin{array}{cccc}
 1 & 0 & 0 & 0 \\
0 & -1 & 2 \mp 2 & 0 \\
0 & 2 \pm 2 & -1 & 0 \\
0 & 0 & 0 & 1
\end{array}\right],
\end{equation}
and the parametrization of the entangling surface SL$(2,\mathbb{R})$ group element 
\begin{equation}
h = \left[\begin{array}{cc} a & b \\ c  & d\end{array}\right],
\end{equation}
the Sklyanin bracket \eqref{eq: b,b poisson} is explicitly given by
\begin{equation}
\left[\begin{array}{cccc}
\{a,a\} & \{a,b\} & \{b,a\} & \{b,b\} \\
\{a,c\} & \{a,d\} & \{b,c\} & \{b,d\} \\
\{c,a\} & \{c,b\}  & \{d,a\} & \{d,b\} \\
\{c,c\} & \{c,d\} & \{d,c\} & \{d,d\}
\end{array}\right] = \frac{\pi}{k}\left[\begin{array}{cccc}
0 & ab & -ab & 0 \\
ac & 2bc & 0 & bd \\
-ac & 0  & -2bc & -bd \\
0 & cd & -cd & 0
\end{array}\right],
\end{equation}
leading to
\begin{gather}
\label{eq:posal}
\{a,b\} = \frac{\pi}{k}ab, \quad \{a,c\} = \frac{\pi}{k} ac, \quad \{a,d\} = \frac{2\pi}{k} bc, \\
\label{eq:posal2}
\{b,c\} = 0, \quad \{b,d\} = \frac{\pi}{k} bd, \quad \{c,d\} = \frac{\pi}{k} cd.
\end{gather}
We can identify this Poisson algebra \eqref{eq:posal}-\eqref{eq:posal2} as the classical $\hbar \to 0$ limit, in the sense of \eqref{eq: classical limit}, of the coordinate algebra $\mathcal{F}(\tx{SL}_q (2, \mathbb{R}))$ \cite{faddeev1988quantization}, generated by $a,b,c,d$ with
$q = \exp{(\frac{2\pi i}{k-2\hbar})}$:\footnote{The usual shift $k\to k-2\hbar$ by the dual Coxeter number is a quantum effect suppressed by a power of $\hbar$.}
\begin{gather}
ab = q^{\frac{\hbar}{2}} ba, \quad ac=q^{\frac{\hbar}{2}} ca, \quad ad-da=(q^{\frac{\hbar}{2}} -q^{-\frac{\hbar}{2}})bc, \\
bc=cb, \quad bd=q^{\frac{\hbar}{2}} db, \quad cd=q^{\frac{\hbar}{2}} dc.
\end{gather} 
We note that this classical limit should not be confused with the undeformed limit ``$q\to 1$'', which can be taken by letting $k\to +\infty$. In the latter case, we obtain the trivial commuting coordinate algebra $\mathcal{F}(\tx{SL}(2,\mathbb{R}))$ of $\tx{SL}(2,\mathbb{R})$ instead of the Poisson algebra described by the relations \eqref{eq:posal}-\eqref{eq:posal2}. 
Finally, the quantum determinant condition $ad-q^{\frac{\hbar}{2}} bc=1$ reduces to $ad-bc=1$ in this limit. The function $ad-bc$ can be indeed easily checked to be a Casimir function of the above Poisson algebra \eqref{eq:posal}.
To avoid confusion, a scheme of the different limits is as follows:

\begin{figure}
    \centering
    \includegraphics[width=0.7\linewidth]{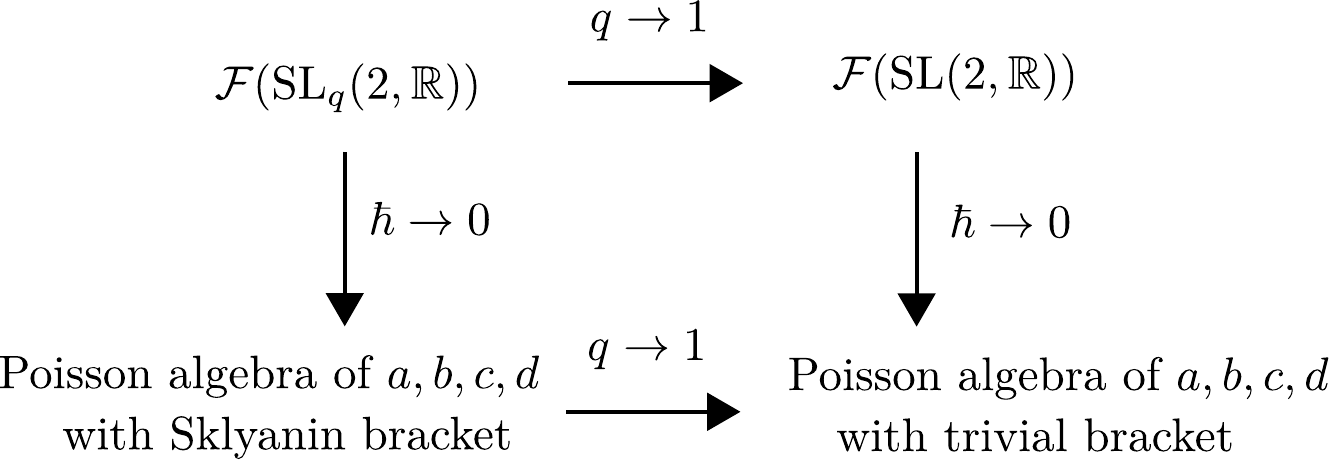}
    \caption{Two different notions of ``classical limits"}
\end{figure}

\textbf{STS-bracket \eqref{eq: STS bracket}:} \\
We parameterize the dual SL$(2,\mathbb{R})$ Poisson manifold by a judicious choice of coordinate functions $K,J^+,J^-$ as
\begin{equation}
\label{eq:coduapoi}
    m = \left(
\begin{array}{cc}
K^{}-\frac{(2 \pi) ^2}{k^2}J^+J^-  & \frac{2 \pi i }{k} J^- K^{-\frac{1}{2}}\\
 \frac{2 \pi  i }{k}J^+ K^{-\frac{1}{2}}&  K^{-1} \\
\end{array}
\right),
\end{equation}
which automatically satisfy the determinant condition $m_{11}m_{22}-m_{12}m_{21}=1$ as a constraint on the algebra. The two coordinates $J^\pm$ are chosen purely imaginary in this description. With this coordinatization we obtain from the STS bracket \eqref{eq: STS bracket} the following algebra relations:
\begin{align}
\label{eq:coalgcl}
\{K,J^+\} = \frac{2\pi}{k}J^+ K, \qquad
\{K,J^-\} = -\frac{2\pi}{k}J^- K,  \qquad
\{J^+,J^-\} = -\frac{k}{2\pi}(K-K^{-1}), 
\end{align}
which were written down in the past in \cite{Gawedzki:1990jc}. These relations also appeared recently in \cite{Blommaert:2023opb,Blommaert:2023wad} in the context of the classical symmetry algebra of the $q$-Schwarzian model describing double-scaled SYK.

Just as above with the Sklyanin bracket, these relations should be viewed as the classical $\hbar \to 0$ version of the defining quantum algebra relations. Indeed, the U$_q(\mathfrak{sl}(2,\mathbb{R}))$ quantum algebra is generated by the relations \cite{Faddeev:1987ih}
\begin{align}
K J^+ = q^{\hbar} J^+ K, \qquad
K J^- = q^{-\hbar} J^-K,  \qquad
[J^+,J^-] = \hbar^2 \frac{K-K^{-1}}{q^{\frac{\hbar}{2}}-q^{-\frac{\hbar}{2}}}, 
\end{align}
Setting $q = \exp{(\frac{2\pi i}{k-2\hbar})}$, and letting $\hbar \to 0$ as in \eqref{eq: classical limit} again, we obtain the leading relations:
\begin{align}
[K, J^+] \approx  i\hbar\frac{2\pi}{k} J^+ K, \qquad
[K, J^-] \approx - i\hbar\frac{2\pi}{k} J^- K,  \qquad
[J^+,J^-] \approx - i\hbar \frac{k}{2\pi}(K-K^{-1}), 
\end{align}
which are precisely the Poisson algebra relations of \eqref{eq:coalgcl}. Note that $\text{Tr}(m)$ is explicitly computed from \eqref{eq:coduapoi} and is indeed the Casimir function of this Poisson algebra. \\

\textbf{Dressing transformation brackets \eqref{eq: m, b dressing transform}:} \\
Finally, the dressing brackets are explicitly evaluated into ($K=q^{H}$)
\begin{align}
\label{eq:dres1}
\{K,h_{ij}\} &=  \frac{2\pi}{k} (h T_H)_{ij}K, \quad \text{ or } \quad \{H,h_{ij}\} = -i (h T_H)_{ij}\,, \\
\label{eq:dres2}
\{J^-K^{-\frac{1}{2}},h_{ij}\} &=  -i (h T_-)_{ij}K^{-1}, \\
\label{eq:dres3}
\{J^+K^{-\frac{1}{2}},h_{ij}\} &= -i (h T_+)_{ij}K^{-1},
\end{align}
with the explicit generators in the fundamental representation:
\begin{equation}
T_H = \frac{1}{2}\left[\begin{array}{cc} 1 & 0 \\ 0  & -1\end{array}\right], \quad 
T_+ = \left[\begin{array}{cc} 0 & 1 \\ 0  & 0\end{array}\right], \quad  T_- = \left[\begin{array}{cc} 0 & 0 \\ 1  & 0\end{array}\right],
\end{equation}
allowing us to interpret the RHS of \eqref{eq:dres1}-\eqref{eq:dres3} as the action of $m$ on the $h$ boundary group element.\footnote{The explicit factors of $i$ are consistent with the fact that our coordinatization \eqref{eq:coduapoi} required that all of $H,J^+$ and $J^-$ are purely imaginary.} In coordinates, the resulting relations \eqref{eq:dres2} and \eqref{eq:dres3} are deformed with the Cartan generator $K$ as compared to the linear case. 

The linear limit $q\to 1^-$ of all three brackets leads to the relations:
\begin{gather}
\{H,J^+\}=-iJ^+, \quad \{H,J^-\}=iJ^-, \quad \{J^+,J^-\} = -2iH, \\
\{H,h_{ij}\} = -i (hT_H)_{ij}, \quad \{J^+,h_{ij}\} = -i(hT_+)_{ij}, \quad \{J^-,h_{ij}\} = -i (hT_-)_{ij}, \\
\{h_{ij},h_{kl}\} = 0,
\end{gather}
the Sklyanin bracket (the last line) becoming trivial. This limit is the known non-abelian edge state algebra shown in the introduction for non-abelian gauge theories (see eqns \eqref{eq:1chalg}, \eqref{eq:3chalg} and \eqref{eq:2chalg} for the three real classical charges $iH, iJ^+$ and $iJ^-$).

\subsection{Incompleteness of the One-sided Poisson Algebra and Causality}

We have encountered an incompleteness in the Poisson algebra for the one-sided system in Section~\ref{sec: poisson algebra of sqrt CS}. 
The completion of the Poisson algebra precisely encodes the $r$-matrix, which is directly visible in the boundary WZNW algebra \eqref{eq: g, g poison r}. From the WZNW boundary perspective, this Poisson algebra incompleteness reflects an apparent violation of causality in chiral models. 
To illustrate this and the origin of the incompleteness itself, in this subsection,\footnote{
We set $G= \tx{U}(1)$ for simplicity and use Dirac's method to derive the Poisson bracket. As a technical comment for experts, this may not be directly applicable if $G$ is nonabelian for the following reason. The Wess-Zumino term is topological and thus does not contribute to the Hamiltonian. However, adding the WZ term changes the equation of motion, which implies that the associated symplectic structure has to change. Due to the presence of the WZ term, one needs to use the multisymplectic formalism to derive the symplectic form for the general WZNW model \cite{Kijowski:1973gi, Kijowski:1976ze, Gawedzki:1990jc}.} we explicitly illustrate this in the chiral U(1) WZNW model with the Floreanini-Jackiw action \cite{Floreanini:1987as}:
\begin{equation}
S_{\text{FJ}} = \frac{k}{4\pi}\int dt dx \left(\partial_t \phi \partial_x \phi - (\partial_x \phi)^2\right).
\end{equation}
With the conjugate momentum
\begin{equation}
\pi_\phi(x,t) = \frac{k}{4\pi}\partial_x \phi(x,t),
\end{equation}
the equal time canonical Poisson bracket is \cite{Sonnenschein:1988ug}: 
\begin{equation}
\label{eq:phipi}
\{\phi(x,t),\pi_\phi(y,t)\} = \frac{1}{2}\delta(x-y). 
\end{equation}
The factor $1/2$
is well-known for the chiral boson system, and corresponds to the fact that $\pi_\phi = \frac{k}{4\pi}\partial_x \phi$ acts as a second-class constraint on the phase space. Integrating the above relation, we get
\begin{equation}
\label{eq:phiphi}
\{\phi(x,t),\phi(y,t)\} = \frac{2\pi}{k} \left(r + \frac{1}{2}\text{sgn}(x-y)\right), 
\end{equation}
where we added the integration constant $r$, which is playing a crucial role in our work in the non-abelian case, see Appendix \ref{app:details} for some details. Here antisymmetry of $r$ forces $r=0$, but this is no longer true in the non-abelian case. This bracket matches with \eqref{eq: g, g poison r} when setting $g(x) = e^{\phi(x)}$ and $C = 1$.
\footnote{One can readily include the winding numbers when $\vert x_{12}\vert > 2\pi$ in \eqref{eq: g, g poison r}, by periodically adding delta-functions to \eqref{eq:phipi} when $x,y$ are living on $S^1$.} The appearance of this a priori undetermined integration constant $r$ in the elementary Poisson brackets can be appreciated by realizing that the chiral WZNW Lagrangians are not Lorentz invariant, and hence the equal time bracket of fields \eqref{eq:phiphi} does not vanish for $x\neq y$ automatically. A relativistic field theory on the other hand would not allow for such an integration constant, since microcausality ($\{\phi(x,t),\phi(y,t)\}=0$ if $x-y$ spacelike) fixes the brackets on a Cauchy slice.

\section{Surface Symmetry Group $G_s$}
\label{s:prop2}

As we described in Eq. \eqref{eq:gauges} and \eqref{eq: Poisson-Lie action}, the one-sided system (when embedded into the two-sided system) is invariant under right multiplication by a group element. 

The right multiplication symmetry group $G$ \eqref{eq:gauges} has the structure of a Poisson manifold as follows \cite{Semenov-Tian-Shansky:1985mgd,Alekseev:1990vr,Falceto:1992bf}. The right multiplication map $\mu$
\begin{equation}
\label{eq: right multiplication}
\mu: P_\odot  \times G \to P_\odot, \qquad (W(x),h) \overset{\mu}{\to} W(x)\cdot h
\end{equation}
is a Poisson-Lie symmetry of the system.\footnote{Viewing the monodromy as an element in the dual Poisson-Lie group $G^*$, it is the moment map associated with the Poisson-Lie symmetry \eqref{eq: right multiplication} \cite{Falceto:1992bf}. In the case of $G=\tx{SL}(2,\mathbb{R})$, the dual is usually denoted as $G^* = \left\{\left(\left[\begin{array}{cc} a & b \\ 0  & 1/a\end{array}\right], \left[\begin{array}{cc} 1/a & 0 \\ c  & a\end{array}\right]\right), a>0,\, b,c \in \mathbb{R}\right\} \simeq \tx{SB}(2,\mathbb{C})$ \cite{grammaticos2004integrability}.} This means it is a Poisson map, which means that the pull-back map $\mu^*$ satisfies
\begin{equation}
\{\mu^*f,\mu^* \tilde{f}\}_{M \times G} = \mu^* \{f,\tilde{f}\}_M, \qquad f,\tilde{f} \in C^\infty(M).
\end{equation}
We apply this to the basis functions $f=W_1(x)\cdot h_1$ and $\tilde{f}=W_2(y)\cdot h_2$, with $M=P_\odot$.
The left-hand side now becomes
\begin{equation}
\{(W_1(x), h_1), (W_2(y), h_2)\} = \{W_1(x),W_2(y)\}h \otimes h + W(x)\otimes W(y) \{h_1,h_2\},
\end{equation}
since this is the Poisson bracket of the product manifold $P_\odot \times G$. This can now be evaluated directly as a Poisson bracket on $P_\odot$ as:
\begin{equation}
\{W_1(x)\cdot h_1, W_2(y) \cdot h_2\} = \frac{2\pi}{k} (W(x)\otimes W(y)) \cdot  (h \otimes h) \, r^{\pm}.
\end{equation}
This finally leads to
\begin{equation}
\frac{2\pi}{k} (W(x)\otimes W(y)) \cdot  (h \otimes h) \, r^{\pm} = \frac{2\pi}{k} (W(x)\otimes W(y))r^\pm (h \otimes h) + W(x)\otimes W(y) \{h_1,h_2\},
\end{equation}
or
\begin{equation}
\label{eq:skly2}
\{h_1,h_2\} = \frac{2\pi}{k} \left[h \otimes h,r\right].
\end{equation}
This is the same Sklyanin bracket as earlier for the coordinate space Poisson algebra (e.g. \eqref{eq:posal} in the case of SL$(2,\mathbb{R})$), except here living on a different manifold (the group $G$ instead of the phase space of the model $P_\odot$).

We conclude that the surface symmetry group $G_s$ = Poisson-Lie group equipped with the Sklyanin bracket \eqref{eq:skly2}.

\section{Comparison of Factorizations}
\label{s:prop3}
The story we presented up to this point is one particular way to proceed and factorize the two-boundary state space in a geometrically natural and minimal way. In this section, we attempt to understand the bigger picture and provide at least a partial classification of all possible factorizations. These options will be unified in the language of ``ungauging'' large gauge transformations at the entangling surface, and hence making them physical. Precisely which and how these candidate degrees of freedom are made physical directly leads to the various possibilities.

We will start by ungauging as few degrees of freedom as possible, and then work our way upwards from there.
We will see that demanding both completeness of the edge degrees of freedom (i.e. allowing a regluing or a surjective gluing map), and demanding the factorization procedure only adds the minimal amount of edge degrees of freedom, actually leads to an almost unique factorization map. The only ambiguity left is $1:1$ with a choice of classical (constant antisymmetric) $r$-matrix.

Our discussion will be summarized into the following table:

\begin{table}[h]
\centering
\begin{tabular}{ |c|c|c| } 
\hline
 & Gluing map & Minimal? \\
 \hline
 Cartan subalgebra \ref{s:fac1} & {\color{red}No} & {\color{red}Subminimal} \\
 Poisson-Lie \ref{s:fac2} & {\color{darkgreen}Yes} & {\color{darkgreen}Yes} \\
 Kac-Moody \ref{s:fac3} & {\color{darkgreen}Yes} & {\color{red}No} \\
 \hline
\end{tabular}
\caption{Three schemes of gluing maps}
\end{table}

\subsection{Cartan Subalgebra ``Factorization''}
\label{s:fac1}
In terms of ungauging degrees of freedom, the first option is to not make any additional would-be-gauge degree of freedom physical. Recall that the group element $W(x^+,x^-) = W(x^+)W^{-1}(x^-)$
with monodromy relation $W(x^++2\pi) = W(x^+) m_+$, $W(x^-+2\pi) = W(x^-)m_-$ satisfies the equivalence relation \eqref{eq:gauges}:
\begin{equation}
\label{eq:gauges2}
W(x^+) \sim W(x^+)h, \qquad W(x^-) \sim W(x^-)h, \qquad m_\pm \sim h^{-1}m_\pm h.
\end{equation}
As emphasized, this is a gauge redundancy for the two-sided description of the system in terms of $(W(x^+),W(x^-),m_\pm)$. This redundancy can be partially gauge-fixed by restricting $m$ to be a representative of a conjugacy class in a maximal torus $T$, a strategy well-studied in the early literature \cite{faddeev1990exchange,Alekseev:1990vr,Chu:1991pn,Balog:1999ep}. Crucially, we have chosen (in this subsection only!) to treat this symmetry as a gauge redundancy also for the one-sided system. 

Proceeding with the construction in this way, there is a trivial (i.e. abelian) Poisson algebra found from the matrix elements of $m$:
\begin{equation}
\{t_i,t_j\} = 0, \qquad i =1 ... \text{rank }G.
\end{equation}
There are hence rank $G$ distinct abelian edge charges.
Quantizing this Poisson algebra is trivial and leads to a state space spanned by the states
\begin{equation}
\label{eq:Hcart}
\mathcal{H} = \{\ket{q_1,... q_{\text{rank }G}}\}.
\end{equation}
This works best for the case of a complex Lie group or real compact Lie group $G$, where there is a single set of conjugacy classes, and $m$ can be taken as an element of a maximal torus $T = U(1)^{\text{rank }G} \subset G$. Each $q_i$ is a discretized abelian charge that can be chosen $\in \mathbb{Z}$ by suitable normalization. For the general case of non-compact real forms, one has to deal with multiple distinct conjugacy classes, leading to sectors where some of the charges $q_i \in \mathbb{R}$.\footnote{In the case of 3d $\Lambda<0$ gravity, which is chirally described by the Teichm\"uller component of SL$(2,\mathbb{R})$ (or SL$^+(2,\mathbb{R}$)), one only needs to consider the (non-compact) hyperbolic conjugacy class, and the resulting single charge would be $\in \mathbb{R}$.}
Equation \eqref{eq:Hcart} is a smaller edge state space than the earlier one constructed, because we decided not to ``ungauge'' the symmetry \eqref{eq:gauges2}.

There is a residual gauge group consisting of all $h\in T$, the same maximal torus. This means the right multiplication Poisson-Lie symmetry is reduced to only the same ``Cartan subgroup''. E.g. in the case of SL$(2,\mathbb{R})$, this would be just $a$ and $d=1/a$ (using that $ad-bc$ is a Casimir function of this Poisson algebra), with vanishing Poisson bracket $\{a,d\}=0$, which is a consistent truncation of the Poisson bracket algebra given above.

The entire Poisson-Lie structure of this choice is trivial: all Poisson brackets are zero, and everything is abelian.\footnote{We note that there is an even more trivial choice, where instead we just turn off the monodromy completely: $m=\mathbf{1}$. This would remove the last edge degree of freedom as well. This option is not reachable however within the language of ungauging a ``large'' gauge symmetry, and is hence of less interest.}

Whereas this leads to a well-defined one-sided system, it is not complete as a factorized edge system.
This system cannot be glued back to the two-sided system, and does not provide a surjective gluing map.

\subsection{Poisson-Lie Factorization}
\label{s:fac2}
As a second option, and our main story in this work, we can physicalize the full group element $h$ at the puncture/entangling surface. Under a mild assumption (the integration constant $r$ being independent of the dynamical variables), we argued this leads to a minimal factorization of the model. We have reduced this minimal factorization to classifying all constant antisymmetric solutions to the MCYBE \eqref{eq: MCYBE}. There is a complete classification of such solutions by Belavin and Drinfel'd \cite{belavin1982solutions} for the case of simple Lie algebras. In the particular case of most interest, SL$(2,\mathbb{R})$, there is only one such solution (up to automorphisms), as explicitly written down above in section \ref{s:sl2}. Higher rank groups have more solutions, and hence a priori physically distinct and consistent minimal edge algebra factorizations. 

3d pure $\Lambda <0$ gravity in Lorentzian signature is governed by the SL$(2,\mathbb{R})\otimes  \text{SL}(2,\mathbb{R})$ isometry group. Relatedly it is described by two Chern-Simons theories based on these simple Lie groups \cite{Achucarro:1986uwr,Witten:1988hc}. With the full 6-dimensional semi-simple Lie algebra $\mathfrak{sl}(2,\mathbb{R})\oplus \mathfrak{sl}(2,\mathbb{R})$, various classes of solutions for a consistent classical $r$-matrix have been explored in the literature, see e.g. \cite{Meusburger:2007ad,Schroers:2011wn,Osei:2017ybk}, non-linearizing this algebra. However, from holography we have an additional strong constraint as follows. The outer holographic boundary Kac-Moody algebras are the symmetry algebras of a boundary CFT. These two Kac-Moody algebras at either boundary represent the chiral and anti-chiral symmetry algebras. In order to respect this chiral factorization at the holographic boundary, we need to have a chirally factorized symmetry algebra also at the entangling surface. This means we have to restrict to a non-linear deformation of $\mathfrak{sl}(2,\mathbb{R})\oplus \mathfrak{sl}(2,\mathbb{R})$ that preserves the factorization into two 3-dimensional algebras. By the above Belavin-Drinfel'd result, the non-linear (minimal) deformation is hence unique also for $\Lambda<0$ 3d Lorentzian gravity.\footnote{For other choices of the cosmological constant $\Lambda$, we leave a more detailed analysis to future work.}

We remark that our analysis was purely classical, and one can ask whether quantization further restricts the minimal extensions (or $r$-matrices).

\subsection{Kac-Moody Factorization}
\label{s:fac3}
\begin{figure}[h]
    \centering
    \includegraphics[width=0.45\linewidth]{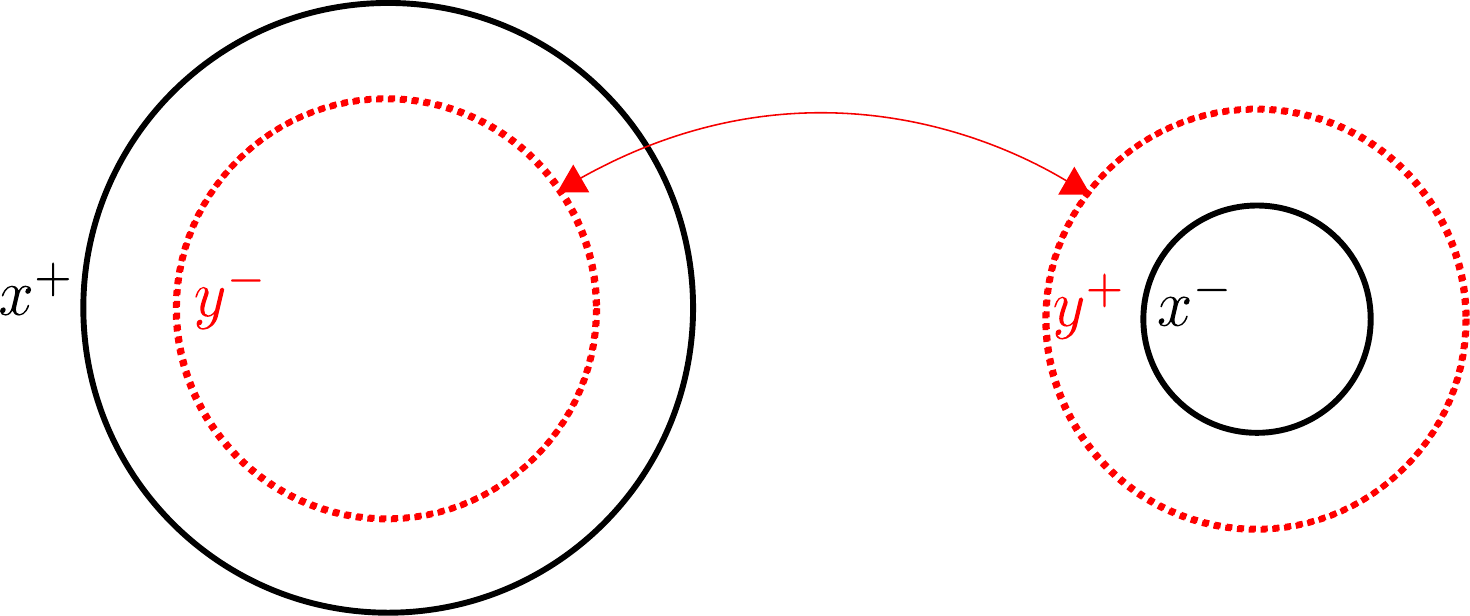}
    \caption{Cutting an annulus with a Kac-Moody algebra of edge states on either side of the cut. This is the Kac-Moody factorization map. The red dashed circle (entangling surface) is geometrically identified between both figures.}
    \label{fig:km}
\end{figure}

Besides the Poisson-Lie gluing map \eqref{eq: gluing morphism}, we can also obtain an annulus by gluing two (two-sided) annuli,
\begin{alignat}
{4}
\; && \mathcal{P}_{ \circledcirc}  \times \mathcal{P}_{ \circledcirc} && \;\twoheadrightarrow\; & \mathcal{P}_\circledcirc,
\label{eq: KM extension}\\
    && \big( W(x^+, y^-) , W(x^-, y^+) \big)\vert_{y^+ = y^-}  \,\,  && \;\mapsto\; & \,\, W(x^+, y^-)  W^{-1}(x^-, y^+)\vert_{y^+ = y^-}.
\end{alignat}
This means we put a chiral WZNW model also at the entangling surface, and consider its degrees of freedom as edge states. The two factors on the left-hand side of the gluing morphism \eqref{eq: KM extension} correspond to the inner and outer parts of an annulus illustrated in Fig. \ref{fig:km}.
The labels $y^\pm$ are coordinates of the Kac-Moody edge modes on the entangling boundary corresponding to the red dashed circle in Fig. \ref{fig:km}. The map \eqref{eq: KM extension} is the Kac-Moody gluing map. Each factor in $\mathcal{P}_{ \circledcirc}  \times \mathcal{P}_{ \circledcirc}$ is equipped with the same Poisson algebra as Eq. \eqref{eq: W, W Poisson KM}, i.e.
\begin{align}
\label{eq: W, W poisson, KM, entangling}
    \{ {W}_1(x^+_1,y^-_1) , {W}_2(x^+_2,y^-_2) \} &= \frac{2 \pi}{k} {W}_1(x^+_1,y^-_1) {W}_2(x^+_2,y^-_2) r^{\tx{KM}}_{12} (y^-_{12}, x^+_{12}), \\
    \{ {W}_1(x^-_1,y^+_1) , {W}_2(x^-_2,y^+_2) \} &= \frac{2 \pi}{k} {W}_1(x^-_1,y^+_1) {W}_2(x^-_2,y^+_2) r^{\tx{KM}}_{12} (y^+_{12}, x^-_{12}).
\end{align}
This type of Kac-Moody factorization is the factorization map for Chern-Simons theory studied extensively in the past directly in the Lagrangian and symplectic structure, see e.g. \cite{Geiller:2017xad} and references therein. Note that $r^{\tx{KM}}_{12}(y,x)$ is only sensitive to topological information, i.e. the winding number and oriented intersection number.

\section{Comments on quantization}
\label{s:quant}

Although this is not the focus of this paper, quantization of the edge mode Heisenberg double algebra \eqref{eq: STS bracket}, \eqref{eq: b,b poisson}, \eqref{eq: m, b dressing transform} is straightforward \cite{semenov1992poisson}. We define the classical $r(x_{12})$-matrix as the linear term in the series expansion of the quantum $R$-matrix as:
\begin{equation}
\label{eq: r to R}
    R(x_{12}) = 1+\frac{2 \pi i\hbar}{k + h^{\vee} \hbar} r(x_{12}) + O(\hbar^2)
\end{equation}
with $h^{\vee}$ the dual Coxeter number. Quantization proceeds by deforming \eqref{eq: W, W poisson, r} into the braiding algebra:
\begin{equation}
\label{eq:qex}
W_1(x_1)W_2(x_2) = W_2(x_2)W_1(x_1)R(x_{12}).
\end{equation}
Associativity for $W_1(x_1)W_2(x_2)W_3(x_3)$ then implies the $R$-matrices $R(x_{12})$ need to satisfy the quantum Yang-Baxter equation:
\begin{equation}
\label{eq: QYBE}
R_{12}(x_{12})R_{13}(x_{13})R_{23}(x_{23}) = R_{23}(x_{23})R_{13}(x_{13})R_{12}(x_{12}).
\end{equation}
Restricting to $\vert x_{12}\vert <2\pi$ (no winding), we denote the $R$-matrices corresponding to the two orderings as $R_\pm$, related to $r^\pm$ through $R_\pm = 1+\frac{2 \pi i\hbar}{k + h^{\vee} \hbar} r^\pm + O(\hbar^2)$,
which satisfy the classical Yang-Baxter equation \eqref{eq:CYBE}. 
Using solely this braiding algebra \eqref{eq:qex}, the above three Poisson brackets \eqref{eq: STS bracket}, \eqref{eq: b,b poisson}, \eqref{eq: m, b dressing transform} are now implied to be consistently quantized to become the exchange relations \cite{semenov1992poisson}:\footnote{A useful intermediate result is $m_1R_+W_2^{-1}(x_2) = R_-W_2^{-1}(x_2)m_1$.}
\begin{align}
\label{eq:qgroup}
R_\pm h_1h_2  &= h_2h_1 R_\pm, \\
\label{eq:qalg}
m_1R_+ m_2 R_-^{-1} &= R_+m_2R_-^{-1} m_1, \\
\label{eq:qdres}
m_1h_2 &= h_2 R_-^{-1} m_1 R_+.
\end{align}
The first two relations \eqref{eq:qgroup}, \eqref{eq:qalg} are just the defining non-commutativity relations for the quantum group $G_q$ and quantum algebra $U_q(\mathfrak{g})$ respectively.
The quantized Heisenberg double is then the operator algebra obtained by taking the product of $G_q$ and $U_q(\mathfrak{g})$, modulo the above exchange relations (in particular including the third one \eqref{eq:qdres}). This defines the phase space operator algebra. Combining in this way the quantum group $G_q$ (e.g. SL$_q(2,\mathbb{R})$) and the quantum algebra $U_q(\mathfrak{g})$ (e.g. $U_q(\mathfrak{sl}(2,\mathbb{R}))$) is natural from a quantum mechanical description in phase space. Indeed, this is simply the non-linear generalization of the observation that a general Lie group element $e^{ix^a T_a}$ ($a=1\hdots \text{dim }G$) is just like $e^{ipx}$ (for a particle on the real line) where the coordinates on the group manifold $x^a$ are conjugate to the generators $T_a$. If one wants to describe physical wavefunctions of the system, we can then choose to work in either a ``momentum'' or a ``position'' basis. Both of these subalgebras are non-commutative in the non-linear case. One describes a physical state most conveniently in the representation basis as $\ket{R,ab}$ ($a,b=1 \hdots \text{dim }R$) by constructing a complete set of states of unitary irrep $R$ matrix elements (i.e. the Peter-Weyl theorem). The conjugate group basis $\ket{g}$ can be used in the linear case, but does not work here as straightforwardly since the coordinates on the quantum group manifold are non-commutative.

Upon quantization, the traced monodromy Casimir function $\text{Tr}(m)$, commuting with the entries of $m$, is replaced by the quantum traced monodromy $\text{Tr}_q(m)$ \cite{Reshetikhin:1990sq} as the Casimir operator of $U_q(\mathfrak{g})$.  This is a deformed trace that involves inserting in the ordinary trace the so-called Drinfel'd element $D$ as: 
\begin{equation}
\text{Tr}_q(\cdot) \equiv \text{Tr}(D\,\cdot).
\end{equation}
For the case of SL$_q(2,\mathbb{R})$, we have $D = q^{\frac{H}{2}}$, with $H$ the Cartan generator of $\mathfrak{sl}(2,\mathbb{R})$.

\section{Concluding Remarks}
We have seen in this work that one can factorize Chern-Simons gauge theory across an entangling surface by introducing quantum group edge degrees of freedom.
Our main result is an identification of the Drinfel'd double Poisson algebra, familiar in the context of classical integrability:
\begin{align}
\label{eq:fin1}
     \{{m}_1, {m}_2  \} &= \frac{2 \pi}{k} (r^+_{12} {m}_1 {m}_2 - {m}_1 r^+_{12} {m}_2 - {m}_2 r^-_{12} {m}_1 +  {m}_1 {m}_2 r^-_{12}), \\
     \{ h_1, h_2 \} &= \frac{2 \pi}{k} [h_1h_2, r], \\
     \label{eq:fin3}
     \{ m_1, h_2 \}
    &= \frac{2 \pi}{k}  h_2\, ({m}_1 r^+_{12} - r^-_{12} {m}_1 ),
\end{align}
where $m$ and $h$ are matrix-valued functions in the Poisson algebra, as a non-linear generalization of the non-abelian linear edge state algebra:
\begin{align}
\{Q_a,Q_b\} &= f_{ab}{}^c Q_c, \\
\{h_1,h_2\} &= 0, \\
\{Q_a,h\} &= h T_a,
\end{align}
one encounters when factorizing gauge field theories across an entangling surface. The surface charge algebra $\{Q_a,Q_b\}$ is to be compared to the monodromy algebra $\{{m}_1, {m}_2  \}$, and the large gauge transformation algebra $\{h_1,h_2\}$ (which is trivial in the linear case) becomes a non-linear algebra. Our non-linear edge algebra \eqref{eq:fin1}-\eqref{eq:fin3} is applicable for a minimal factorization in the case of topological gauge theories, and has been made explicit throughout this work for Chern-Simons theories. \\

Let us close with some speculation that we largely leave for future work. \\

\textbf{Factorization in compact vs non-compact groups} \\
Our factorization map in terms of Poisson-Lie symmetries crucially relied on the classification of all solutions to the split MCYBE \eqref{eq: MCYBE}. As is well-known, the number of such solutions is very dependent on precisely which real form of a complex algebra one considers. For instance, for compact Lie algebras, there is no solution at all \cite{cahen1994some}!
For the same reason, the Riemann-Hilbert factorization problem, where we split $g=g_+(g_-)^{-1}$ in opposing Borel subgroups has no solution for compact groups.
This led the authors of \cite{Gawedzki:1990jc,Alekseev:1990vr,Falceto:1992bf} to always state that the current framework only makes sense upon complexification of the phase space, something we deem unphysical from our perspective of finding physical edge states and factorization. Hence, we believe the factorization procedure generically only works for the split real (or normal) form. We note that precisely this issue motivated past work into more generic ways of making sense of chiral WZNW models in terms of Poisson-Lie groupoids \cite{balog2000chiral}. \\

\textbf{Towards an action principle} \\
In order to define a dynamical system, one needs to specify both its phase space structure and its Hamiltonian. In this work, we have only specified the phase space (Poisson algebra) of the edge states. To give dynamics to these states, one needs to write down a Hamiltonian as well. Various candidate model Hamiltonians that exhibit the Poisson-Lie symmetry exhibited here as a classical Lagrangian symmetry can be found in the literature in various contexts: a ``squashed'' sigma model was written down in \cite{Kawaguchi:2011pf}, a boundary phase space Lagrangian for a particle on SL$_q(2,\mathbb{R})$ was written down in \cite{Blommaert:2023opb,Blommaert:2023wad} (see also \cite{Chu:1994hm}), providing in turn a boundary dual to the 2d bulk Poisson-sigma model which contains the same non-linear Poisson-Lie symmetry algebra \cite{Schaller:1994es,Ikeda:1993aj,Ikeda:1993fh,Cattaneo:2001bp,Grumiller:2003ad}.

However, ultimately edge states live on an entangling surface, which is an infinite redshift surface or black hole horizon according to the one-sided observer. This means in practice that the Hamiltonian is redshifted to zero, and there is no dynamics of the edge states: they are frozen on the horizon \cite{Blommaert:2018oue}. This means the edge theory is quite generically expected to be a purely topological sector, whose sole importance is in its counting of degrees of freedom. In this sense, providing the Poisson structure of the model is all one needs to do to define a complete edge state sector. \\

\textbf{Entanglement and anyonic entropy}  \\
These edge states are viewed according to the one-sided observer as having support only on the entangling surface. Their one-sided energy is zero, due to effectively infinitely redshifting to the entangling surface. As such, only their counting is physical. In the quantum theory, for a given irrep label $j$, one hence automatically finds a contribution to a suitably $q$-deformed von Neumann entropy (see e.g. \cite{Couvreur_2017}):
\begin{equation}
S_{\text{vN}}^q \equiv - \text{Tr}_q(\rho \log \rho) =  \log \dim_q j,
\end{equation}
sometimes called the anyonic entanglement entropy \cite{Bonderson:2017osr}. On the left the quantum trace function is used that is by definition invariant under the adjoint action of the coordinate Hopf algebra SL$_q(2,\mathbb{R})$.

All of this makes the contribution to the entanglement entropy match with the anyon defect entropy. For a quantum state specified with a distribution $P(j)$ of spin labels $j$, the final entropy formula is then
\begin{equation}
S =  \sum_j P(j) \log \dim_q j,
\end{equation}
which is manifestly positive and finite. See also \cite{Delcamp:2016eya}. \\

\textbf{Application to 3d $\Lambda < 0$ gravity} \\
Our initial motivation for this work was to try to understand the observation that factorizing $\Lambda <0$ 3d gravity models seem to require only quantum group edge sector degrees of freedom \cite{Mertens:2022ujr,Wong:2022eiu}. The argument relied on demanding a match between entanglement entropy and thermal entropy of the one-sided observer.
The group theoretical structure of 3d gravity is governed by the Virasoro algebra, based on the quantum group $\text{SL}^+_{q}(2,\mathbb{R}) \times \text{SL}^+_{q}(2,\mathbb{R})$, in parallel to its decomposition into two Chern-Simons theories based on the SL$(2,\mathbb{R})$ Lie group. The $^+$ superscripts here signal a restriction to positive quantum group elements as required to map SL$(2,\mathbb{R})$ into gravity. This specification of the relevant quantum group is familiar in the language of Teichm\"uller or Virasoro TQFT \cite{Verlinde:1989ua,EllegaardAndersen:2011vps,Collier:2023fwi}. In the group theoretical framework, the gravitational entropy of BTZ black holes is found as a defect anyon entropy as
\begin{align}\label{Sbh}
    S_{\text{def}}= \log (\dim_{q}p_{+} \dim_{q}p_{-}) = \log (S_{0}{}^{p_{+}}S_{0}{}^{p_{-}}),
\end{align}
where the $p_\pm$ labels the continuous series representations of two copies of SL$_q(2,\mathbb{R})$,\footnote{These are singled out as self-dual representations, which form a basis of ``functions on the quantum group manifold'' of the modular double $U_q(\mathfrak{sl}(2,\mathbb{R})) \otimes U_{\tilde{q}}(\mathfrak{sl}(2,\mathbb{R}))$ \cite{Faddeev:1999fe,Ponsot:1999uf,Ponsot:2000mt,Kharchev:2001rs,Bytsko:2002br,Bytsko:2006ut}.} which encode the mass $M$ and angular momentum $J$ of a rotating black hole through $M = p_+^2+p_-^2$ and $J=p_+^2-p_-^2$. This identification of BTZ black hole entropy as topological entanglement entropy was first proposed in \cite{McGough:2013gka}. 
Since the relevant group is the split real form of SL$(2)$, our methods apply and the above observations simply mean that 3d gravity should be factorized precisely in the way we have laid out throughout this work. This procedure would lead to a \emph{unique} minimal factorization map as we discussed above in section \ref{s:fac2}. The argument can in principle be generalized to higher spin gravity based on the split real form of SL$(N)$, again allowing us to apply our methods. We find it remarkable that gravity conspires to precisely make this factorization map work.

Holography in 3d gravity plays a big role in constraining the deformation we presented in this work. This was utilized in two places. Firstly, chiral factorization of the boundary CFT implies that the symmetry algebra at the entangling surface is similarly factorized. We expect this is then also true for closed universes (without a holographic boundary), since one could dynamically emit or absorb a baby universe, and the consistency of such an interaction then seems to require the same chiral factorization. Secondly, our minimal factorization map (and not the WNZW factorization) is implied to match the thermal BTZ entropy as an entanglement entropy across the black hole horizon as was explained in detail in \cite{Mertens:2022ujr}.

\section*{Acknowledgments}
We thank A. Belaey, A. Blommaert, L-Y. Hung, J. Papalini, J. Sim\'on, T. Tappeiner and G. Wong for discussions, and J. Sim\'on and G. Wong for comments on our manuscript. We acknowledge financial support from the European Research Council (grant BHHQG-101040024). Funded by the European Union. Views and opinions expressed are however those of the author(s) only and do not necessarily reflect those of the European Union or the European Research Council. Neither the European Union nor the granting authority can be held responsible for them.

\appendix

\section{Poisson Bracket of Wilson Lines}
\label{app: KM double}
We calculate the Poisson bracket of two Wilson lines in $\mathcal{P}_\circledcirc$ by the Kac-Moody algebra double \eqref{eq: KM sym tensor +}, \eqref{eq: KM sym tensor -} as follows. We first write out the path-ordered exponential in infinitesimal segments, and use the Leibniz rule for each segment as:
\begin{align}
    \{ {W}_1(x^+_1,x^-_1) , {W}_2(x^+_2,x^-_2) \}
    &=
    \{ \overleftarrow{P} \exp ( i \int_{x^-_1}^{x^+_1} A_1 ), {W}_2(x^+_2,x^-_2) \}
    \\
    &=
    \int_{x^-_1}^{x^+_1} dy\, W_1 (x^+_1, y) \{ 1+i A_1(y), {W}_2(x^+_2,x^-_2) \} W_1 ( y , x^-_1)
    \end{align}
Applying the Kac-Moody algebra double \eqref{eq: KM sym tensor +}, \eqref{eq: KM sym tensor -} and locality, we then write
\begin{align}  
&\,\,\,\quad \int_{x^-_1}^{x^+_1} dy\, W_1 (x^+_1, y) \{ 1+i A_1(y), {W}_2(x^+_2,x^-_2) \} W_1 ( y , x^-_1)
\\
    &=
    \frac{2\pi }{k} \int_{x^-_1}^{x^+_1} dy\, W_1 (x^+_1, y) [  {C}_{12}{W}_2({x_2^+, x^-_2}) \overunderset{\infty}{n=-\infty}{\sum} \delta (y - x_2^+ - 2\pi n) ] W_1 ( y , x^-_1) \nonumber
    \\
    & \qquad -
    \frac{2\pi }{k} \int_{x^-_1}^{x^+_1} dy\, W_1 (x^+_1, y) [  {W}_2({x_2^+, x^-_2}) {C}_{12} \overunderset{\infty}{n=-\infty}{\sum} \delta (y - x_2^- - 2\pi n) ] W_1 ( y , x^-_1) \nonumber
\end{align}
which can be further worked out as
\begin{align}
    &=
    \frac{2\pi }{k} \overunderset{\infty}{n=-\infty}{\sum}  W_1 (x^+_1, x^+_2 + 2 n \pi)  {C}_{12}{W}_2({x_2^+, x^-_2})   W_1 ( x^+_2 + 2 n \pi , x^-_1) \int_{x^-_1}^{x^+_1} dy\,\delta (y - x_2^+ - 2\pi n) \nonumber
    \\
    &-
    \frac{2\pi }{k} \overunderset{\infty}{n=-\infty}{\sum}  W_1 (x^+_1, x^-_2 + 2n \pi)  {W}_2({x_2^+, x^-_2}) {C}_{12}    W_1 ( x^-_2 + 2n \pi , x^-_1) \int_{x^-_1}^{x^+_1} dy\, \delta (y - x_2^- - 2\pi n) \nonumber
    \\
    &=
    \frac{\pi }{k} \overunderset{\infty}{n=-\infty}{\sum}  W_1 (x^+_1, x^+_2 + 2 n \pi)  {C}_{12}{W}_2({x_2^+, x^-_2})   W_1 ( x^+_2 + 2 n \pi , x^-_1) \tx{sgn}(x^+_{12}-2n \pi) \nonumber
    \\
    &-
    \frac{\pi }{k} \overunderset{\infty}{n=-\infty}{\sum}  W_1 (x^+_1, x^-_2 + 2n \pi) {W}_2({x_2^+, x^-_2}) {C}_{12}    W_1 ( x^-_2 + 2n \pi , x^-_1) \tx{sgn}(x^-_{12}- 2n \pi).
\end{align}
We can apply the flatness of the connection to see
\begin{align}
    &\quad \,\,\,W_1 (x^+_1, x^+_2 + 2 n \pi)  {C}_{12}{W}_2({x_2^+, x^-_2})   W_1 ( x^+_2 + 2 n \pi , x^-_1)
    \\
    &=
    W_1 ( x^+_1, x^-_1 ) W_2 ( x^+_2, x^-_2 ) W_1 (x^-_1, x^+_2 + 2 n \pi)  W_2^{-1} ( x^+_2, x^-_2 ){C}_{12}{W}_2({x_2^+, x^-_2})   W_1 ( x^+_2 + 2 n \pi , x^-_1) \nonumber
    \\
    &=
    W_1 ( x^+_1, x^-_1 ) W_2 ( x^+_2, x^-_2 ) W_1 (x^-_1, x^+_2 + 2 n \pi)  W_1 ( x^+_2, x^-_2 ){C}_{12}{W}_1({x_2^-, x^+_2})   W_1 ( x^+_2 + 2 n \pi , x^-_1) \nonumber
    \\
    &=
    W_1 ( x^+_1, x^-_1 ) W_2 ( x^+_2, x^-_2 ) W_1 (x^-_1, x^+_2 + 2 n \pi)  W_1 ( x^+_2, x^-_2 ){C}_{12}{W}_1({x_2^-, x^+_2})   W_1 ( x^+_2  , x^-_1-2n \pi) \nonumber
    \\
    &=
    W_1 ( x^+_1, x^-_1 ) W_2 ( x^+_2, x^-_2 ) W_1 (x^-_1 -2n \pi, x^+_2 )  W_1 ( x^+_2, x^-_2 ){C}_{12}{W}_1(x_2^- , x^-_1-2n \pi) \nonumber
    \\
    &=
    W_1 ( x^+_1, x^-_1 ) W_2 ( x^+_2, x^-_2 ) W_1 (x^-_1 -2n \pi, x^-_2 ){C}_{12}{W}_1(x_2^- , x^-_1-2n \pi). \nonumber
\end{align}
Similarly,
\begin{align}
    &\quad\,\,\,W_1 (x^+_1, x^-_2 + 2n \pi) {W}_2({x_2^+, x^-_2}) {C}_{12}    W_1 ( x^-_2 + 2n \pi , x^-_1)
    \\
    &=
    W_1 ( x^+_1, x^-_1 ) W_2 ( x^+_2, x^-_2 ) W_1 (x^-_1, x^+_1) W_1 (x^+_1, x^-_2 + 2n \pi)  {C}_{12}    W_1 ( x^-_2 + 2n \pi , x^-_1) \nonumber
    \\
    &=
    W_1 ( x^+_1, x^-_1 ) W_2 ( x^+_2, x^-_2 ) W_1 (x^-_1, x^-_2 + 2n \pi)  {C}_{12}    W_1 ( x^-_2 + 2n \pi , x^-_1) \nonumber
    \\
    &=
    W_1 ( x^+_1, x^-_1 ) W_2 ( x^+_2, x^-_2 ) W_1 (x^-_1-2n \pi, x^-_2)  {C}_{12}    W_1 ( x^-_2 , x^-_1 - 2n \pi). \nonumber
\end{align}
Combining all equations above, we derive the Poisson bracket \eqref{eq: W, W Poisson KM}. By taking derivatives, it is easy to see the Poisson bracket \eqref{eq: W, W Poisson KM} implies the Kac-Moody algebra double \eqref{eq: KM sym tensor +}, \eqref{eq: KM sym tensor -}.

\section{Complementary Symplectic Form Perspective}
\label{app:symlform}
 The left-chiral WZNW model has the symplectic form \cite{Gawedzki:1990jc,Falceto:1992bf}:
\begin{equation}
\label{eq:gaw}
\Omega_L = \frac{k}{4\pi} \int_0^{2\pi}dx \text{Tr}\left[W^{-1} \delta W \wedge \partial_x (W^{-1}\delta W)\right] + \frac{k}{4\pi} \text{Tr}\big[W^{-1}\delta W(0) \wedge \delta m_+m_+^{-1}\big] - \frac{k}{4\pi} \rho(m_+),
\end{equation}
where we added a correction two-form $- \frac{k}{4\pi} \rho(m_+)$, compared to a naive splitting of the symplectic form of the non-chiral one. To simplify notation, we from here on drop the ${}_+$ subscript of the monodromy variable $m$. This $\Omega_L$ is also precisely the symplectic form of Chern-Simons theory for the degrees of freedom living only on $S^1_L$ (again up to the addition of the two-form $- \frac{k}{4\pi} \rho(m)$) as we show in Appendix \ref{app:symplCS}.

Requiring that this symplectic form is closed, requires
\begin{equation}
\delta \Omega_L = \frac{k}{12\pi} \text{Tr}\left[(m^{-1}\delta m) \wedge (m^{-1}\delta m) \wedge (m^{-1}\delta m)\right] - \frac{k}{4\pi}  \delta \rho(m) \overset{!}{=}0,
\end{equation}
then leads to a constraint of the functional form of $\rho$. The solution space of this constraint has a deep relation with classical integrability as follows. A classical (constant antisymmetric) $r$-matrix encodes a unique decomposition of the algebra in Borel subalgebras as $\mathfrak{g}=\mathfrak{g}_+-\mathfrak{g}_-$ and group $G \to (G^+,G^-)$ with $g=g^+(g^{-})^{-1}$. If one now sets 
\begin{equation}
\rho(m) = \text{Tr}\left[(m^-)^{-1}\delta m^- \wedge (m^+)^{-1}\delta m^+\right],
\end{equation}
using this decomposition for the monodromy element $m=m^+(m^-)^{-1}$, then we readily get $\delta \Omega_L = 0$.\footnote{In doing this calculation, note that $\text{Tr}\left[((m^+)^{-1}\delta m^+) \wedge ((m^+)^{-1}\delta m^+) \wedge ((m^+)^{-1}\delta m^+)\right] =0$, and also for the term with all minus superscripts, by Cartan's solvability criterion for Borel subalgebras.} It was stated and partially shown in \cite{Gawedzki:1990jc,Falceto:1992bf}, and fully proven later in \cite{balog2000chiral}, that any choice of additive term $\rho$ in the symplectic form is $1:1$ with an antisymmetric constant solution of the MCYBE \eqref{eq: MCYBE}, with that $r$-matrix precisely the one determining the decomposition $m=m^-(m^{+})^{-1}$. So we reach precisely the same conclusion: the factorization map is fully determined by a choice of classical $r$-matrix.

As the main example, the ``standard'' classical $r$-matrix (corresponding to the quasi-triangular Hopf algebra), for Lie algebra $\mathfrak{g}$ with Chevalley basis generators $e_\alpha$, is given by 
\begin{equation}
r = \frac{1}{4}\sum_{\alpha >0} (e_\alpha \otimes e_{-\alpha} - e_{-\alpha} \otimes e_{\alpha}),
\end{equation}
and corresponds to the splitting $g \to (g^-,g^+)$ into the negative and positive Borel subgroups as $(n^-t,n^+t^{-1})$, restricted such that the Cartan subgroup elements are each others' inverse. Note that this requires strictly speaking that we work either with the complexified Lie group, or the maximally noncompact (i.e. split real) form.

The symplectic form of the two-sided theory is $\Omega = \Omega_L + \Omega_R$ and satisfies $\delta \Omega=0$, requiring one picks the same $\rho$ on both sides (up to a sign). The $R$-theory hence has just $\rho \to -\rho$ or swaps the $+$ and $-$ sectors in the decomposition. This is equivalent to the Cayley transform $r_{12}^\pm \to r_{12}^{\mp}$ as we found above.

This discussion might seem a bit ad hoc, since one just introduces the two-form $\rho$ by hand in \eqref{eq:gaw},\footnote{The two-form $\rho$ is not globally defined in phase space, since the Wess-Zumino term $\text{Tr}[(m^{-1}\delta m)^3]$ is not an exact form.} and then shows it is intimately related with integrability. In our approach in the previous subsections, we showed that the emergence of a Poisson-Lie group is inevitable. On the other hand, the idea of modifying the symplectic structure to facilitate edge states and factorization is well appreciated in the literature, starting with the work \cite{Donnelly:2016auv}.

\section{Symplectic Form in Chern-Simons Theory}
\label{app:symplCS}

\begin{figure}[h]
    \centering
    \includegraphics[width=0.22\linewidth]{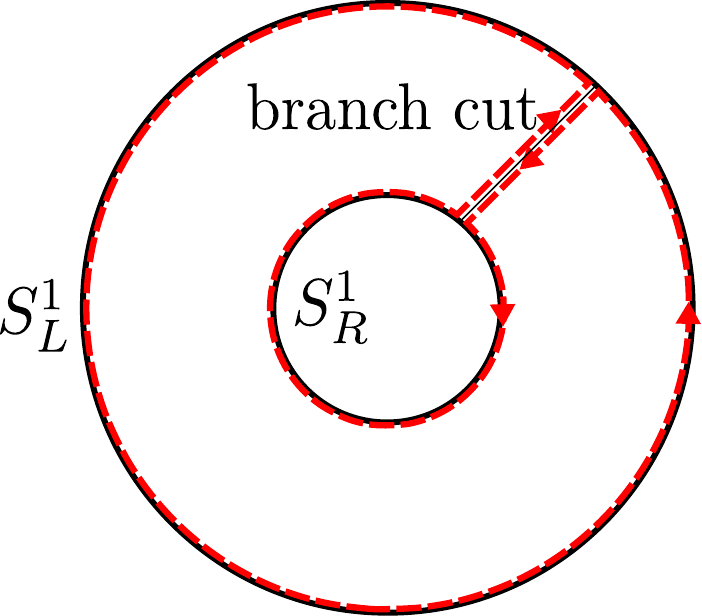}
    \caption{Annulus with arbitrary radial cut depicted for the function $W(x)$, such that the function is single-valued within this region.}
    \label{fig:cut}
\end{figure}

In this section, we show that the symplectic form of Chern-Simons theory \cite{Atiyah:1982fa,Alekseev:1993rj} 
matches that of the two chiral components of the WZNW model. We denote by $d$ the exterior derivative on spacetime and $\delta$ the exterior derivative on the phase space. These two exterior derivatives commute $\delta d  = d \delta$.\footnote{To compare with the symplectic form of the full non-chiral WZNW model derived in \cite{Gawedzki:1990jc}, the multisymplectic formalism of \cite{Kijowski:1973gi, Kijowski:1976ze} is useful where,
\begin{equation}
    \delta d  + d \delta = 0.
\end{equation}
Instead, we follow the convention of commuting exterior derivatives.
}
Varying the Chern-Simons Lagrangian \eqref{eq: bulk chern-simons lagrangian},
\begin{equation}
    \delta L = -\frac{k}{2 \pi } \tr ( \delta A F ) \underbrace{-\frac{k}{4 \pi} d \tr   ( A \delta A )}_{ = d \theta},
\end{equation}
we can read off the symplectic potential density,
\begin{equation}
    \theta = -\frac{k}{4 \pi} \tr   ( A \delta A ).
\end{equation}
The symplectic current $\omega$ is the variation of the known symplectic potential density:
\begin{equation}
    \omega = \delta \theta = -\frac{k}{4 \pi} \tr   ( \delta A  \wedge \delta A ).
\end{equation}
Inserting $A = -i dW W^{-1}$, we have
\begin{equation}
    \omega = \frac{k}{4 \pi} \tr   ( \delta (dW W^{-1}) \wedge \delta (dW W^{-1} )).
\end{equation}
To proceed, we will utilize the following lemma repeatedly:
\begin{equation}
\label{eq:lemma}
\delta (dW W^{-1}) = W d(W^{-1}\delta W) W^{-1}.
\end{equation}
This leads to the symplectic form on the annulus $\circledcirc$:
\begin{equation}
\Omega = \int_{\circledcirc} \omega =  \frac{k}{4 \pi} \int_{\circledcirc} \tr   (d(W^{-1}\delta W)  \wedge  d(W^{-1}\delta W)) =  \frac{k}{4 \pi} \int_{\circledcirc} d \tr   (W^{-1}\delta W  \wedge  d(W^{-1}\delta W)),
\end{equation}
reducing by Stokes' theorem to a boundary contribution from the outer resp. inner circles:
\begin{equation}
\frac{k}{4 \pi}\oint_{S^1_L}  \tr   (W^{-1}\delta W  \wedge  d(W^{-1}\delta W)) + \frac{k}{4 \pi}\oint_{S^1_R}  \tr   (W^{-1}\delta W  \wedge  d(W^{-1}\delta W)).
\end{equation}
There is a further boundary contribution coming from the non-trivial monodromy, which one can visualize as two ``radial'' segments of the boundary contour, separated by a $2\pi$ rotation (see Fig. \ref{fig:cut}). In fact any line connecting the inner and outer circle works.
These radial segments lead to the contribution:
\begin{equation}
\frac{k}{4 \pi}\int_{\text{branch cut}} \left[ \tr   (W^{-1}\delta W  \wedge  d(W^{-1}\delta W))\vert_{2\pi} - \tr   (W^{-1}\delta W  \wedge  d(W^{-1}\delta W))\vert_{0} \right].
\end{equation}
To evaluate these, we introduce the monodromy variable as
\begin{equation}
m \equiv W^{-1}|_{x} \,W|_{x+2\pi}, \qquad dm=0,
\end{equation}
which is the same matrix everywhere on the annulus ($dm=0$). We again use \eqref{eq:lemma} to obtain
\begin{align}
    \tr [ \delta W W^{-1} \wedge  \delta( dW W^{-1} ) ]\vert^{2\pi}_0 
    &=
    \tr [ \delta m m^{-1} \wedge W^{-1} \delta (dW W^{-1}) W ]\vert_0
    \\
    &=
    \tr [ \delta m m^{-1}\wedge d (W^{-1} \delta W ) ]\vert_0
    \\
    &=
    d \tr [ \delta m m^{-1} \wedge W^{-1} \delta W ]\vert_0,
\end{align}
to finally evaluate this contribution as a boundary contribution (coming from two points, one on the inner and one on the outer circle) as:
\begin{equation}
\frac{k}{4 \pi} \tr [ \delta m m^{-1} \wedge W^{-1} \delta W ]\vert_0.
\end{equation}
Collecting all terms, we find in the end:
\begin{align}
\label{eq: symplectic form CS}
\Omega = \int_{\circledcirc} \omega &= \frac{k}{4 \pi}\oint_{S^1_L}  \tr   (W^{-1}\delta W  \wedge  d(W^{-1}\delta W)) + \frac{k}{4 \pi} \tr [ \delta m_+ m_+^{-1} \wedge W^{-1} \delta W ]\vert_0 \\
&+ \frac{k}{4 \pi}\oint_{S^1_R}  \tr   (W^{-1}\delta W  \wedge  d(W^{-1}\delta W)) - \frac{k}{4 \pi} \tr [ \delta m_-^{-1} m_- \wedge W^{-1} \delta W ]\vert_0. \nonumber
\end{align}

The choice of coordinate $0$ is arbitrary, and $\Omega$ can be rewritten in the same way with any reference coordinate. The symplectic form \eqref{eq: symplectic form CS} is precisely the symplectic form of the WZNW model obtained in \cite{Gawedzki:1990jc}, and split in this way into its left- and right-chiral contributions.

\section{Comments on Affine Poisson Structures}
\label{app:comments}
In Section \ref{s:poqg} we have defined the Poisson structure of the local variables $g(x)$, choosing to match the $r$-matrix of the non-local $W(x)$ Wilson lines to define the physically most sensible model. More precisely, we embed the minimal extended phase space $\mathcal{P}_\odot$ into a larger space $\{(g(x),h)\}$. In doing so, we must define the Poisson structure on that larger phase space. Suppose
\begin{equation}
\label{eq: g,g poison, r'}
    \{ {g}_{1}(x_1), {g}_{2}(x_2) \} = \frac{2 \pi}{k} {g}_{1}(x_1) {g}_{2}(x_2) \left(r'_{12} + \frac{1}{2}  \,  {C}_{12}\, \tx{sgn}(x_{12})\right), \quad |x_{12}|< 2\pi ,
\end{equation}
with $r'$ not necessarily equal to $r$. Combining Eq.\eqref{eq: W, W poisson, r}, \eqref{eq: W=gb}, \eqref{eq: locality} and \eqref{eq: g,g poison, r'}, we have
\begin{equation}
\label{eq: affine Pisson}
    \{ h_1, h_2 \} = \frac{2\pi}{k} (h_1 h_2 r_{12} - r'_{12} h_1 h_2).
\end{equation}
The Poisson bracket \eqref{eq: affine Pisson} still defines a compatible Poisson structure on $G$, which is called the affine Poisson structure~\cite{Semenov-Tian-Shansky:1985mgd, weinstein1990affine}. We can see that embedding the minimal extension into the phase space of $\{g(x),h\}$ introduces additional ambiguities besides the original classical $r$-matrix. Now the larger extension is labeled by two solutions of MCYBE \eqref{eq: MCYBE}, $r$ and $r'$. The Drinfel'd double $DG$ of the Poisson-Lie group $G$ equipped with the Poisson bracket $\eqref{eq: affine Pisson}$ reduces to the cotangent bundle $T^*G$ of $G$ in the classical limit $k \to \infty$. Thus, all the Drinfel'd doubles labeled by $(r,r')$ are sensible deformations of the same cotangent bundle $T^*G$. However, some of these deformations have degenerate Poisson structures. If $r'=r$, then the corresponding Poisson structure is almost non-degenerate except on a measure-zero subset in the phase space. If $DG$ with $r' = r$ is homeomorphic as a manifold to $G \times G^*$, then $DG$ is a symplectic manifold~\cite{Semenov-Tian-Shansky:1985mgd, Alekseev:1993qs}.

\section{Some details on chiral boson constrained system}
\label{app:details}
In equations (6) and (7) of \cite{Sonnenschein:1988ug}, the Dirac brackets were written down with the phase space constraint $\rho(x) \equiv \pi_\phi - \frac{k}{4\pi}\partial_x \phi = 0$. To find the Dirac brackets, one needs the inverse of the constraint matrix $\{\rho(x),\rho(y)\} = -2 \frac{k}{4\pi} \delta'(x-y)$, which requires solving 
\begin{equation}
-2\int dy \delta'(x-y)f(y,z) = \frac{4\pi}{k}\delta(x-z), \quad -2\int dy f(x,y)\delta'(y-z) = \frac{4\pi}{k}\delta(x-z),
\end{equation}
or 
\begin{equation}
\partial_y f(y,z) = -\frac{1}{2} \frac{4\pi}{k} \delta(y-z), \quad \partial_z f(y,z) = +\frac{1}{2} \frac{4\pi}{k} \delta(y-z),
\end{equation}
solved by
\begin{equation}
f(y,z) = -\frac{1}{4} \frac{4\pi}{k} \text{sgn}(y-z) - \frac{2\pi}{k}r,
\end{equation}
which allows for an integration constant that we denoted as $- \frac{2\pi}{k}r$, corresponding to the fact that the constraint matrix has an eigenvector with eigenvalue zero. This then leads to an additive constant in the Dirac bracket \eqref{eq:phiphi}.

\bibliographystyle{ourbst}
\bibliography{ref.bib}

\providecommand{\href}[2]{#2}\begingroup\raggedright\begin{thebibliography}{100}

\bibitem{Ryu:2006bv}
S.~Ryu and T.~Takayanagi, ``{Holographic derivation of entanglement entropy from AdS/CFT},'' \href{http://dx.doi.org/10.1103/PhysRevLett.96.181602}{{\em Phys. Rev. Lett.} {\bfseries 96} (2006) 181602}, \href{http://arxiv.org/abs/hep-th/0603001}{{\ttfamily arXiv:hep-th/0603001}}.

\bibitem{VanRaamsdonk:2010pw}
M.~Van~Raamsdonk, ``{Building up spacetime with quantum entanglement},'' \href{http://dx.doi.org/10.1142/S0218271810018529}{{\em Gen. Rel. Grav.} {\bfseries 42} (2010) 2323--2329}, \href{http://arxiv.org/abs/1005.3035}{{\ttfamily arXiv:1005.3035 [hep-th]}}.

\bibitem{Donnelly:2011hn}
W.~Donnelly, ``{Decomposition of entanglement entropy in lattice gauge theory},'' \href{http://dx.doi.org/10.1103/PhysRevD.85.085004}{{\em Phys. Rev. D} {\bfseries 85} (2012) 085004}, \href{http://arxiv.org/abs/1109.0036}{{\ttfamily arXiv:1109.0036 [hep-th]}}.

\bibitem{Buividovich:2008gq}
P.~V. Buividovich and M.~I. Polikarpov, ``{Entanglement entropy in gauge theories and the holographic principle for electric strings},'' \href{http://dx.doi.org/10.1016/j.physletb.2008.10.032}{{\em Phys. Lett. B} {\bfseries 670} (2008) 141--145}, \href{http://arxiv.org/abs/0806.3376}{{\ttfamily arXiv:0806.3376 [hep-th]}}.

\bibitem{Casini:2013rba}
H.~Casini, M.~Huerta, and J.~A. Rosabal, ``{Remarks on entanglement entropy for gauge fields},'' \href{http://dx.doi.org/10.1103/PhysRevD.89.085012}{{\em Phys. Rev. D} {\bfseries 89} no.~8, (2014) 085012}, \href{http://arxiv.org/abs/1312.1183}{{\ttfamily arXiv:1312.1183 [hep-th]}}.

\bibitem{Donnelly:2014gva}
W.~Donnelly, ``{Entanglement entropy and nonabelian gauge symmetry},'' \href{http://dx.doi.org/10.1088/0264-9381/31/21/214003}{{\em Class. Quant. Grav.} {\bfseries 31} no.~21, (2014) 214003}, \href{http://arxiv.org/abs/1406.7304}{{\ttfamily arXiv:1406.7304 [hep-th]}}.

\bibitem{Lin:2018bud}
J.~Lin and D.~Radi\v{c}evi\'c, ``{Comments on defining entanglement entropy},'' \href{http://dx.doi.org/10.1016/j.nuclphysb.2020.115118}{{\em Nucl. Phys. B} {\bfseries 958} (2020) 115118}, \href{http://arxiv.org/abs/1808.05939}{{\ttfamily arXiv:1808.05939 [hep-th]}}.

\bibitem{Ghosh:2015iwa}
S.~Ghosh, R.~M. Soni, and S.~P. Trivedi, ``{On The Entanglement Entropy For Gauge Theories},'' \href{http://dx.doi.org/10.1007/JHEP09(2015)069}{{\em JHEP} {\bfseries 09} (2015) 069}, \href{http://arxiv.org/abs/1501.02593}{{\ttfamily arXiv:1501.02593 [hep-th]}}.

\bibitem{Donnelly:2015hxa}
W.~Donnelly and A.~C. Wall, ``{Geometric entropy and edge modes of the electromagnetic field},'' \href{http://dx.doi.org/10.1103/PhysRevD.94.104053}{{\em Phys. Rev. D} {\bfseries 94} no.~10, (2016) 104053}, \href{http://arxiv.org/abs/1506.05792}{{\ttfamily arXiv:1506.05792 [hep-th]}}.

\bibitem{Blommaert:2018rsf}
A.~Blommaert, T.~G. Mertens, H.~Verschelde, and V.~I. Zakharov, ``{Edge State Quantization: Vector Fields in Rindler},'' \href{http://dx.doi.org/10.1007/JHEP08(2018)196}{{\em JHEP} {\bfseries 08} (2018) 196}, \href{http://arxiv.org/abs/1801.09910}{{\ttfamily arXiv:1801.09910 [hep-th]}}.

\bibitem{Blommaert:2018oue}
A.~Blommaert, T.~G. Mertens, and H.~Verschelde, ``{Edge dynamics from the path integral \textemdash{} Maxwell and Yang-Mills},'' \href{http://dx.doi.org/10.1007/JHEP11(2018)080}{{\em JHEP} {\bfseries 11} (2018) 080}, \href{http://arxiv.org/abs/1804.07585}{{\ttfamily arXiv:1804.07585 [hep-th]}}.

\bibitem{Geiller:2019bti}
M.~Geiller and P.~Jai-akson, ``{Extended actions, dynamics of edge modes, and entanglement entropy},'' \href{http://dx.doi.org/10.1007/JHEP09(2020)134}{{\em JHEP} {\bfseries 09} (2020) 134}, \href{http://arxiv.org/abs/1912.06025}{{\ttfamily arXiv:1912.06025 [hep-th]}}.

\bibitem{Donnelly:2016jet}
W.~Donnelly and G.~Wong, ``{Entanglement branes in a two-dimensional string theory},'' \href{http://dx.doi.org/10.1007/JHEP09(2017)097}{{\em JHEP} {\bfseries 09} (2017) 097}, \href{http://arxiv.org/abs/1610.01719}{{\ttfamily arXiv:1610.01719 [hep-th]}}.

\bibitem{Donnelly:2018ppr}
W.~Donnelly and G.~Wong, ``{Entanglement branes, modular flow, and extended topological quantum field theory},'' \href{http://dx.doi.org/10.1007/JHEP10(2019)016}{{\em JHEP} {\bfseries 10} (2019) 016}, \href{http://arxiv.org/abs/1811.10785}{{\ttfamily arXiv:1811.10785 [hep-th]}}.

\bibitem{Donnelly:2020teo}
W.~Donnelly, Y.~Jiang, M.~Kim, and G.~Wong, ``{Entanglement entropy and edge modes in topological string theory. Part I. Generalized entropy for closed strings},'' \href{http://dx.doi.org/10.1007/JHEP10(2021)201}{{\em JHEP} {\bfseries 10} (2021) 201}, \href{http://arxiv.org/abs/2010.15737}{{\ttfamily arXiv:2010.15737 [hep-th]}}.

\bibitem{Jiang:2020cqo}
Y.~Jiang, M.~Kim, and G.~Wong, ``{Entanglement entropy and edge modes in topological string theory. Part II. The dual gauge theory story},'' \href{http://dx.doi.org/10.1007/JHEP10(2021)202}{{\em JHEP} {\bfseries 10} (2021) 202}, \href{http://arxiv.org/abs/2012.13397}{{\ttfamily arXiv:2012.13397 [hep-th]}}.

\bibitem{Wen:2016snr}
X.~Wen, S.~Matsuura, and S.~Ryu, ``{Edge theory approach to topological entanglement entropy, mutual information and entanglement negativity in Chern-Simons theories},'' \href{http://dx.doi.org/10.1103/PhysRevB.93.245140}{{\em Phys. Rev. B} {\bfseries 93} no.~24, (2016) 245140}, \href{http://arxiv.org/abs/1603.08534}{{\ttfamily arXiv:1603.08534 [cond-mat.mes-hall]}}.

\bibitem{Wong:2017pdm}
G.~Wong, ``{A note on entanglement edge modes in Chern Simons theory},'' \href{http://dx.doi.org/10.1007/JHEP08(2018)020}{{\em JHEP} {\bfseries 08} (2018) 020}, \href{http://arxiv.org/abs/1706.04666}{{\ttfamily arXiv:1706.04666 [hep-th]}}.

\bibitem{Geiller:2017xad}
M.~Geiller, ``{Edge modes and corner ambiguities in 3d Chern\textendash{}Simons theory and gravity},'' \href{http://dx.doi.org/10.1016/j.nuclphysb.2017.09.010}{{\em Nucl. Phys. B} {\bfseries 924} (2017) 312--365}, \href{http://arxiv.org/abs/1703.04748}{{\ttfamily arXiv:1703.04748 [gr-qc]}}.

\bibitem{Fliss:2020cos}
J.~R. Fliss and R.~G. Leigh, ``{Interfaces and the extended Hilbert space of Chern-Simons theory},'' \href{http://dx.doi.org/10.1007/JHEP07(2020)009}{{\em JHEP} {\bfseries 07} (2020) 009}, \href{http://arxiv.org/abs/2004.05123}{{\ttfamily arXiv:2004.05123 [hep-th]}}.

\bibitem{Klinger:2023qna}
M.~S. Klinger, R.~G. Leigh, and P.-C. Pai, ``{Extended phase space in general gauge theories},'' \href{http://dx.doi.org/10.1016/j.nuclphysb.2023.116404}{{\em Nucl. Phys. B} {\bfseries 998} (2024) 116404}, \href{http://arxiv.org/abs/2303.06786}{{\ttfamily arXiv:2303.06786 [hep-th]}}.

\bibitem{kong2018drinfeld}
L.~Kong and H.~Zheng, ``Drinfeld center of enriched monoidal categories,'' {\em Advances in Mathematics} {\bfseries 323} (2018) 411--426.

\bibitem{kong2024enriched}
L.~Kong, W.~Yuan, Z.-H. Zhang, and H.~Zheng, ``Enriched monoidal categories i: centers,'' {\em Quantum Topology} (2024) .

\bibitem{Mertens:2022ujr}
T.~G. Mertens, J.~Sim\'on, and G.~Wong, ``{A proposal for 3d quantum gravity and its bulk factorization},'' \href{http://dx.doi.org/10.1007/JHEP06(2023)134}{{\em JHEP} {\bfseries 06} (2023) 134}, \href{http://arxiv.org/abs/2210.14196}{{\ttfamily arXiv:2210.14196 [hep-th]}}.

\bibitem{McGough:2013gka}
L.~McGough and H.~Verlinde, ``{Bekenstein-Hawking Entropy as Topological Entanglement Entropy},'' \href{http://dx.doi.org/10.1007/JHEP11(2013)208}{{\em JHEP} {\bfseries 11} (2013) 208}, \href{http://arxiv.org/abs/1308.2342}{{\ttfamily arXiv:1308.2342 [hep-th]}}.

\bibitem{Delcamp:2016eya}
C.~Delcamp, B.~Dittrich, and A.~Riello, ``{On entanglement entropy in non-Abelian lattice gauge theory and 3D quantum gravity},'' \href{http://dx.doi.org/10.1007/JHEP11(2016)102}{{\em JHEP} {\bfseries 11} (2016) 102}, \href{http://arxiv.org/abs/1609.04806}{{\ttfamily arXiv:1609.04806 [hep-th]}}.

\bibitem{Akers:2024wab}
C.~Akers, R.~M. Soni, and A.~Y. Wei, ``{Multipartite edge modes and tensor networks},'' \href{http://dx.doi.org/10.21468/SciPostPhysCore.7.4.070}{{\em SciPost Phys. Core} {\bfseries 7} (2024) 070}, \href{http://arxiv.org/abs/2404.03651}{{\ttfamily arXiv:2404.03651 [hep-th]}}.

\bibitem{Dupuis:2020ndx}
M.~Dupuis, L.~Freidel, F.~Girelli, A.~Osumanu, and J.~Rennert, ``{On the origin of the quantum group symmetry in 3d quantum gravity},'' \href{http://arxiv.org/abs/2006.10105}{{\ttfamily arXiv:2006.10105 [gr-qc]}}.

\bibitem{Chua:2023ios}
W.~Z. Chua and Y.~Jiang, ``{Hartle-Hawking state and its factorization in 3d gravity},'' \href{http://dx.doi.org/10.1007/JHEP03(2024)135}{{\em JHEP} {\bfseries 03} (2024) 135}, \href{http://arxiv.org/abs/2309.05126}{{\ttfamily arXiv:2309.05126 [hep-th]}}.

\bibitem{Chen:2024unp}
L.~Chen, L.-Y. Hung, Y.~Jiang, and B.-X. Lao, ``{Deriving the non-perturbative gravitational dual of quantum Liouville theory from BCFT operator algebra},'' \href{http://dx.doi.org/10.21468/SciPostPhys.19.6.163}{{\em SciPost Phys.} {\bfseries 19} no.~6, (2025) 163}, \href{http://arxiv.org/abs/2403.03179}{{\ttfamily arXiv:2403.03179 [hep-th]}}.

\bibitem{Blommaert:2018iqz}
A.~Blommaert, T.~G. Mertens, and H.~Verschelde, ``{Fine Structure of Jackiw-Teitelboim Quantum Gravity},'' \href{http://dx.doi.org/10.1007/JHEP09(2019)066}{{\em JHEP} {\bfseries 09} (2019) 066}, \href{http://arxiv.org/abs/1812.00918}{{\ttfamily arXiv:1812.00918 [hep-th]}}.

\bibitem{Mertens:2022aou}
T.~G. Mertens, ``{Quantum exponentials for the modular double and applications in gravity models},'' \href{http://dx.doi.org/10.1007/JHEP09(2023)106}{{\em JHEP} {\bfseries 09} (2023) 106}, \href{http://arxiv.org/abs/2212.07696}{{\ttfamily arXiv:2212.07696 [hep-th]}}.

\bibitem{Belaey:2025ijg}
A.~Belaey, T.~G. Mertens, and T.~Tappeiner, ``{Quantum group origins of edge states in double-scaled SYK},'' \href{http://arxiv.org/abs/2503.20691}{{\ttfamily arXiv:2503.20691 [hep-th]}}.

\bibitem{Witten:1989wf}
E.~Witten, ``{Gauge Theories and Integrable Lattice Models},'' \href{http://dx.doi.org/10.1016/0550-3213(89)90232-0}{{\em Nucl. Phys. B} {\bfseries 322} (1989) 629--697}.

\bibitem{Guadagnini:1989th}
E.~Guadagnini, M.~Martellini, and M.~Mintchev, ``{Chern-Simons Holonomies and the Appearance of Quantum Groups},'' \href{http://dx.doi.org/10.1016/0370-2693(90)91963-C}{{\em Phys. Lett. B} {\bfseries 235} (1990) 275--281}.

\bibitem{Guadagnini:1989tj}
E.~Guadagnini, M.~Martellini, and M.~Mintchev, ``{Braids and Quantum Group Symmetry in Chern-Simons Theory},'' \href{http://dx.doi.org/10.1016/0550-3213(90)90443-H}{{\em Nucl. Phys. B} {\bfseries 336} (1990) 581--609}.

\bibitem{Alvarez-Gaume:1988bek}
L.~Alvarez-Gaume, C.~Gomez, and G.~Sierra, ``{Hidden Quantum Symmetries in Rational Conformal Field Theories},'' \href{http://dx.doi.org/10.1016/0550-3213(89)90604-4}{{\em Nucl. Phys. B} {\bfseries 319} (1989) 155--186}.

\bibitem{Alvarez-Gaume:1988izd}
L.~Alvarez-Gaume, C.~Gomez, and G.~Sierra, ``{Quantum Group Interpretation of Some Conformal Field Theories},'' \href{http://dx.doi.org/10.1016/0370-2693(89)90027-0}{{\em Phys. Lett. B} {\bfseries 220} (1989) 142--152}.

\bibitem{Alvarez-Gaume:1989blj}
L.~Alvarez-Gaume, C.~Gomez, and G.~Sierra, ``{Duality and Quantum Groups},'' \href{http://dx.doi.org/10.1016/0550-3213(90)90116-U}{{\em Nucl. Phys. B} {\bfseries 330} (1990) 347--398}.

\bibitem{Gawedzki:1990jc}
K.~Gawedzki, ``{Classical origin of quantum group symmetries in Wess-Zumino-Witten conformal field theory},'' \href{http://dx.doi.org/10.1007/BF02102735}{{\em Commun. Math. Phys.} {\bfseries 139} (1991) 201--214}.

\bibitem{Alekseev:1990vr}
A.~Alekseev and S.~L. Shatashvili, ``{Quantum Groups and {WZW} Models},'' \href{http://dx.doi.org/10.1007/BF02097372}{{\em Commun. Math. Phys.} {\bfseries 133} (1990) 353--368}.

\bibitem{Chu:1991pn}
M.-f. Chu, P.~Goddard, I.~Halliday, D.~I. Olive, and A.~Schwimmer, ``{Quantization of the Wess-Zumino-Witten model on a circle},'' \href{http://dx.doi.org/10.1016/0370-2693(91)90746-D}{{\em Phys. Lett. B} {\bfseries 266} (1991) 71--81}.

\bibitem{Falceto:1992bf}
F.~Falceto and K.~Gawedzki, ``{Lattice Wess-Zumino-Witten model and quantum groups},'' \href{http://dx.doi.org/10.1016/0393-0440(93)90056-K}{{\em J. Geom. Phys.} {\bfseries 11} (1993) 251--279}, \href{http://arxiv.org/abs/hep-th/9209076}{{\ttfamily arXiv:hep-th/9209076}}.

\bibitem{Gaberdiel:1994vv}
M.~R. Gaberdiel, ``{An Explicit construction of the quantum group in chiral WZW models},'' \href{http://dx.doi.org/10.1007/BF02101238}{{\em Commun. Math. Phys.} {\bfseries 173} (1995) 357--378}, \href{http://arxiv.org/abs/hep-th/9407186}{{\ttfamily arXiv:hep-th/9407186}}.

\bibitem{Slingerland:2001ea}
J.~K. Slingerland and F.~A. Bais, ``{Quantum groups and nonAbelian braiding in quantum Hall systems},'' \href{http://dx.doi.org/10.1016/S0550-3213(01)00308-X}{{\em Nucl. Phys. B} {\bfseries 612} (2001) 229--290}, \href{http://arxiv.org/abs/cond-mat/0104035}{{\ttfamily arXiv:cond-mat/0104035}}.

\bibitem{Fock:1998nu}
V.~V. Fock and A.~A. Rosly, ``{Poisson structure on moduli of flat connections on Riemann surfaces and r matrix},'' {\em Am. Math. Soc. Transl.} {\bfseries 191} (1999) 67--86, \href{http://arxiv.org/abs/math/9802054}{{\ttfamily arXiv:math/9802054}}.

\bibitem{Alekseev:1994pa}
A.~Y. Alekseev, H.~Grosse, and V.~Schomerus, ``{Combinatorial quantization of the Hamiltonian Chern-Simons theory},'' \href{http://dx.doi.org/10.1007/BF02099431}{{\em Commun. Math. Phys.} {\bfseries 172} (1995) 317--358}, \href{http://arxiv.org/abs/hep-th/9403066}{{\ttfamily arXiv:hep-th/9403066}}.

\bibitem{Alekseev:1994au}
A.~Y. Alekseev, H.~Grosse, and V.~Schomerus, ``{Combinatorial quantization of the Hamiltonian Chern-Simons theory. 2.},'' \href{http://dx.doi.org/10.1007/BF02101528}{{\em Commun. Math. Phys.} {\bfseries 174} (1995) 561--604}, \href{http://arxiv.org/abs/hep-th/9408097}{{\ttfamily arXiv:hep-th/9408097}}.

\bibitem{Meusburger:2003ta}
C.~Meusburger and B.~J. Schroers, ``{Poisson structure and symmetry in the Chern-Simons formulation of (2+1)-dimensional gravity},'' \href{http://dx.doi.org/10.1088/0264-9381/20/11/318}{{\em Class. Quant. Grav.} {\bfseries 20} (2003) 2193--2234}, \href{http://arxiv.org/abs/gr-qc/0301108}{{\ttfamily arXiv:gr-qc/0301108}}.

\bibitem{Ballesteros:2015twa}
A.~Ballesteros, F.~J. Herranz, and P.~Naranjo, ``{Towards ( 3+1 ) gravity through Drinfel'd doubles with cosmological constant},'' \href{http://dx.doi.org/10.1016/j.physletb.2015.04.041}{{\em Phys. Lett. B} {\bfseries 746} (2015) 37--43}, \href{http://arxiv.org/abs/1502.07518}{{\ttfamily arXiv:1502.07518 [gr-qc]}}.

\bibitem{Meusburger:2008bs}
C.~Meusburger and K.~Noui, ``{The Hilbert space of 3d gravity: quantum group symmetries and observables},'' \href{http://dx.doi.org/10.4310/ATMP.2010.v14.n6.a3}{{\em Adv. Theor. Math. Phys.} {\bfseries 14} no.~6, (2010) 1651--1715}, \href{http://arxiv.org/abs/0809.2875}{{\ttfamily arXiv:0809.2875 [gr-qc]}}.

\bibitem{Freidel:2021ajp}
L.~Freidel, C.~Goeller, and E.~R. Livine, ``{The quantum gravity disk: Discrete current algebra},'' \href{http://dx.doi.org/10.1063/5.0051647}{{\em J. Math. Phys.} {\bfseries 62} no.~10, (2021) 102303}, \href{http://arxiv.org/abs/2103.13171}{{\ttfamily arXiv:2103.13171 [hep-th]}}.

\bibitem{Delcamp:2016yix}
C.~Delcamp, B.~Dittrich, and A.~Riello, ``{Fusion basis for lattice gauge theory and loop quantum gravity},'' \href{http://dx.doi.org/10.1007/JHEP02(2017)061}{{\em JHEP} {\bfseries 02} (2017) 061}, \href{http://arxiv.org/abs/1607.08881}{{\ttfamily arXiv:1607.08881 [hep-th]}}.

\bibitem{Bonzom:2014wva}
V.~Bonzom, M.~Dupuis, F.~Girelli, and E.~R. Livine, ``{Deformed phase space for 3d loop gravity and hyperbolic discrete geometries},'' \href{http://arxiv.org/abs/1402.2323}{{\ttfamily arXiv:1402.2323 [gr-qc]}}.

\bibitem{Dupuis:2017otn}
M.~Dupuis, L.~Freidel, and F.~Girelli, ``{Discretization of 3d gravity in different polarizations},'' \href{http://dx.doi.org/10.1103/PhysRevD.96.086017}{{\em Phys. Rev. D} {\bfseries 96} no.~8, (2017) 086017}, \href{http://arxiv.org/abs/1701.02439}{{\ttfamily arXiv:1701.02439 [gr-qc]}}.

\bibitem{Moore:1989ni}
G.~W. Moore and N.~Reshetikhin, ``{A Comment on Quantum Group Symmetry in Conformal Field Theory},'' \href{http://dx.doi.org/10.1016/0550-3213(89)90219-8}{{\em Nucl. Phys. B} {\bfseries 328} (1989) 557--574}.

\bibitem{Reshetikhin:1991tc}
N.~Reshetikhin and V.~G. Turaev, ``{Invariants of three manifolds via link polynomials and quantum groups},'' \href{http://dx.doi.org/10.1007/BF01239527}{{\em Invent. Math.} {\bfseries 103} (1991) 547--597}.

\bibitem{Donnelly:2016auv}
W.~Donnelly and L.~Freidel, ``{Local subsystems in gauge theory and gravity},'' \href{http://dx.doi.org/10.1007/JHEP09(2016)102}{{\em JHEP} {\bfseries 09} (2016) 102}, \href{http://arxiv.org/abs/1601.04744}{{\ttfamily arXiv:1601.04744 [hep-th]}}.

\bibitem{atiyah1988topological}
M.~F. Atiyah, ``Topological quantum field theory,'' {\em Publications Math{\'e}matiques de l'IH{\'E}S} {\bfseries 68} (1988) 175--186.

\bibitem{Elitzur:1989nr}
S.~Elitzur, G.~W. Moore, A.~Schwimmer, and N.~Seiberg, ``{Remarks on the Canonical Quantization of the Chern-Simons-Witten Theory},'' \href{http://dx.doi.org/10.1016/0550-3213(89)90436-7}{{\em Nucl. Phys. B} {\bfseries 326} (1989) 108--134}.

\bibitem{Babelon:2003qtg}
O.~Babelon, D.~Bernard, and M.~Talon, \href{http://dx.doi.org/10.1017/CBO9780511535024}{{\em {Introduction to Classical Integrable Systems}}}.
\newblock Cambridge Monographs on Mathematical Physics. Cambridge University Press, 2003.

\bibitem{grammaticos2004integrability}
B.~Grammaticos, Y.~Kosmann-Schwarzbach, and K.~Tamizhmani, {\em Integrability of nonlinear systems}.
\newblock Springer, 2004.

\bibitem{Carlip:1998uc}
S.~Carlip, \href{http://dx.doi.org/10.1017/CBO9780511564192}{{\em {Quantum gravity in 2+1 dimensions}}}.
\newblock Cambridge Monographs on Mathematical Physics. Cambridge University Press, 12, 2003.

\bibitem{Goldman1986InvariantFO}
W.~M. Goldman, ``Invariant functions on lie groups and hamiltonian flows of surface group representations,'' \href{https://api.semanticscholar.org/CorpusID:10414720}{{\em Inventiones mathematicae} {\bfseries 85} (1986) 263--302}.

\bibitem{Turaev1991}
V.~G. Turaev, ``Skein quantization of poisson algebras of loops on surfaces,'' {\em Annales scientifiques de l'École Normale Supérieure} {\bfseries 24} no.~6, (1991) 635--704.

\bibitem{Semenov-Tian-Shansky:1985mgd}
M.~A. Semenov-Tian-Shansky, ``{Dressing transformations and Poisson group actions},'' \href{http://dx.doi.org/10.2977/prims/1195178514}{{\em Publ. Res. Inst. Math. Sci. Kyoto} {\bfseries 21} (1985) 1237--1260}.

\bibitem{balog2000chiral}
J.~Balog, L.~Feher, and L.~Palla, ``{Chiral extensions of the WZNW phase space, Poisson-Lie symmetries and groupoids},'' \href{http://dx.doi.org/10.1016/S0550-3213(99)00738-5}{{\em Nucl. Phys. B} {\bfseries 568} (2000) 503--542}, \href{http://arxiv.org/abs/hep-th/9910046}{{\ttfamily arXiv:hep-th/9910046}}.

\bibitem{Alekseev:1991wq}
A.~Alekseev, L.~D. Faddeev, M.~Semenov-Tian-Shansky, and A.~Volkov, ``{The Unraveling of the quantum group structure in the WZNW theory},''.

\bibitem{Alekseev:1992wn}
A.~Alekseev, L.~D. Faddeev, and M.~Semenov-Tian-Shansky, ``{Hidden quantum groups inside Kac-Moody algebra},'' \href{http://dx.doi.org/10.1007/BF02097628}{{\em Commun. Math. Phys.} {\bfseries 149} (1992) 335--345}.

\bibitem{faddeev1990exchange}
L.~D. Faddeev, ``{On the Exchange Matrix for {WZNW} Model},'' \href{http://dx.doi.org/10.1007/BF02278003}{{\em Commun. Math. Phys.} {\bfseries 132} (1990) 131--138}.

\bibitem{faddeev1982integrable}
L.~Faddeev, ``Integrable models in 1+ 1 dimensional quantum field theory,'' tech. rep., CEA Centre d'Etudes Nucleaires de Saclay, 1982.

\bibitem{Han:2025xfl}
M.~Han, ``{Hamiltonian quantization of complex Chern-Simons theory at level-$k$},'' \href{http://arxiv.org/abs/2504.16367}{{\ttfamily arXiv:2504.16367 [hep-th]}}.

\bibitem{Balog:1999wy}
J.~Balog, L.~Feher, and L.~Palla, ``{The Chiral WZNW phase space and its Poisson-Lie groupoid},'' \href{http://dx.doi.org/10.1016/S0370-2693(99)00965-X}{{\em Phys. Lett. B} {\bfseries 463} (1999) 83--92}, \href{http://arxiv.org/abs/hep-th/9907050}{{\ttfamily arXiv:hep-th/9907050}}.

\bibitem{semenov1992poisson}
M.~Semenov-Tyan-Shanskii, ``Poisson-lie groups. the quantum duality principle and the twisted quantum double,'' {\em Theoretical and Mathematical Physics} {\bfseries 93} no.~2, (1992) 1292--1307.

\bibitem{sklyanin1979complete}
E.~K. Sklyanin, ``On complete integrability of the landau-lifshitz equation,'' tech. rep., 1979.

\bibitem{sklyanin1982some}
E.~K. Sklyanin, ``Some algebraic structures connected with the yang--baxter equation,'' {\em Funktsional'nyi Analiz i ego Prilozheniya} {\bfseries 16} no.~4, (1982) 27--34.

\bibitem{sklyanin1983some}
E.~K. Sklyanin, ``{Some algebraic structures connected with the Yang-Baxter equation},'' \href{http://dx.doi.org/10.1007/BF01077848}{{\em Funct. Anal. Appl.} {\bfseries 16} (1982) 263--270}.

\bibitem{Drinfeld:1986in}
V.~G. Drinfeld, ``{Quantum groups},'' \href{http://dx.doi.org/10.1007/BF01247086}{{\em Zap. Nauchn. Semin.} {\bfseries 155} (1986) 18--49}.

\bibitem{Alekseev:1993qs}
A.~Y. Alekseev and A.~Z. Malkin, ``{Symplectic structures associated to Lie-Poisson groups},'' \href{http://dx.doi.org/10.1007/BF02105190}{{\em Commun. Math. Phys.} {\bfseries 162} (1994) 147--174}, \href{http://arxiv.org/abs/hep-th/9303038}{{\ttfamily arXiv:hep-th/9303038}}.

\bibitem{semenov2008integrable}
M.~Semenov-Tian-Shansky, {\em Integrable systems: the r-matrix approach}.
\newblock Research Inst. for Math. Sciences, Kyoto Univ., 2008.

\bibitem{kosmann2007lie}
Y.~Kosmann-Schwarzbach, ``Lie bialgebras, poisson lie groups and dressing transformations,'' in {\em Integrability of Nonlinear Systems: Proceedings of the CIMPA School Pondicherry University, India, 8--26 January 1996}, pp.~104--170.
\newblock Springer, 2007.

\bibitem{Witten:1991we}
E.~Witten, ``{On quantum gauge theories in two-dimensions},'' \href{http://dx.doi.org/10.1007/BF02100009}{{\em Commun. Math. Phys.} {\bfseries 141} (1991) 153--209}.

\bibitem{Witten:1988hc}
E.~Witten, ``{(2+1)-Dimensional Gravity as an Exactly Soluble System},'' \href{http://dx.doi.org/10.1016/0550-3213(88)90143-5}{{\em Nucl. Phys. B} {\bfseries 311} (1988) 46}.

\bibitem{Achucarro:1986uwr}
A.~Achucarro and P.~K. Townsend, ``{A Chern-Simons Action for Three-Dimensional anti-De Sitter Supergravity Theories},'' \href{http://dx.doi.org/10.1016/0370-2693(86)90140-1}{{\em Phys. Lett. B} {\bfseries 180} (1986) 89}.

\bibitem{faddeev1988quantization}
L.~D. Faddeev, N.~Y. Reshetikhin, and L.~A. Takhtajan, ``{Quantization of Lie Groups and Lie Algebras},'' {\em Alg. Anal.} {\bfseries 1} no.~1, (1989) 178--206.

\bibitem{Blommaert:2023opb}
A.~Blommaert, T.~G. Mertens, and S.~Yao, ``{Dynamical actions and q-representation theory for double-scaled SYK},'' \href{http://dx.doi.org/10.1007/JHEP02(2024)067}{{\em JHEP} {\bfseries 02} (2024) 067}, \href{http://arxiv.org/abs/2306.00941}{{\ttfamily arXiv:2306.00941 [hep-th]}}.

\bibitem{Blommaert:2023wad}
A.~Blommaert, T.~G. Mertens, and S.~Yao, ``{The q-Schwarzian and Liouville gravity},'' \href{http://dx.doi.org/10.1007/JHEP11(2024)054}{{\em JHEP} {\bfseries 11} (2024) 054}, \href{http://arxiv.org/abs/2312.00871}{{\ttfamily arXiv:2312.00871 [hep-th]}}.

\bibitem{Faddeev:1987ih}
L.~D. Faddeev, N.~Y. Reshetikhin, and L.~A. Takhtajan, ``{Quantization of Lie Groups and Lie Algebras},'' {\em Alg. Anal.} {\bfseries 1} no.~1, (1989) 178--206.

\bibitem{Kijowski:1973gi}
J.~Kijowski, ``{A finite-dimensional canonical formalism in the classical field theory},'' \href{http://dx.doi.org/10.1007/BF01645975}{{\em Commun. Math. Phys.} {\bfseries 30} (1973) 99--128}.

\bibitem{Kijowski:1976ze}
J.~Kijowski and W.~Szczyrba, ``{A Canonical Structure for Classical Field Theories},'' \href{http://dx.doi.org/10.1007/BF01608496}{{\em Commun. Math. Phys.} {\bfseries 46} (1976) 183--206}.

\bibitem{Floreanini:1987as}
R.~Floreanini and R.~Jackiw, ``{Selfdual Fields as Charge Density Solitons},'' \href{http://dx.doi.org/10.1103/PhysRevLett.59.1873}{{\em Phys. Rev. Lett.} {\bfseries 59} (1987) 1873}.

\bibitem{Sonnenschein:1988ug}
J.~Sonnenschein, ``{Chiral bosons},'' \href{http://dx.doi.org/10.1016/0550-3213(88)90339-2}{{\em Nucl. Phys. B} {\bfseries 309} (1988) 752--770}.

\bibitem{Balog:1999ep}
J.~Balog, L.~Feher, and L.~Palla, ``{Classical Wakimoto realizations of chiral WZNW Bloch waves},'' \href{http://dx.doi.org/10.1088/0305-4470/33/5/310}{{\em J. Phys. A} {\bfseries 33} (2000) 945--956}, \href{http://arxiv.org/abs/hep-th/9910112}{{\ttfamily arXiv:hep-th/9910112}}.

\bibitem{belavin1982solutions}
A.~A. Belavin and V.~G. Drinfeld, ``Solutions of the classical yang--baxter equation for simple lie algebras,'' {\em Funktsional'nyi Analiz i ego Prilozheniya} {\bfseries 16} no.~3, (1982) 1--29.

\bibitem{Meusburger:2007ad}
C.~Meusburger and B.~J. Schroers, ``{Quaternionic and Poisson-Lie structures in 3d gravity: The Cosmological constant as deformation parameter},'' \href{http://dx.doi.org/10.1063/1.2973040}{{\em J. Math. Phys.} {\bfseries 49} (2008) 083510}, \href{http://arxiv.org/abs/0708.1507}{{\ttfamily arXiv:0708.1507 [gr-qc]}}.

\bibitem{Schroers:2011wn}
B.~J. Schroers, ``{Quantum gravity and non-commutative spacetimes in three dimensions: a unified approach},'' \href{http://dx.doi.org/10.5506/APhysPolBSupp.4.379}{{\em Acta Phys. Polon. Supp.} {\bfseries 4} (2011) 379--402}, \href{http://arxiv.org/abs/1105.3945}{{\ttfamily arXiv:1105.3945 [gr-qc]}}.

\bibitem{Osei:2017ybk}
P.~K. Osei and B.~J. Schroers, ``{Classical r-matrices for the generalised Chern{\textendash}Simons formulation of 3d gravity},'' \href{http://dx.doi.org/10.1088/1361-6382/aaaa5e}{{\em Class. Quant. Grav.} {\bfseries 35} no.~7, (2018) 075006}, \href{http://arxiv.org/abs/1708.07650}{{\ttfamily arXiv:1708.07650 [hep-th]}}.

\bibitem{Reshetikhin:1990sq}
N.~Y. Reshetikhin and M.~A. Semenov-Tian-Shansky, ``{Central extensions of quantum current groups},'' \href{http://dx.doi.org/10.1007/BF01045884}{{\em Lett. Math. Phys.} {\bfseries 19} (1990) 133--142}.

\bibitem{cahen1994some}
M.~Cahen, S.~Gutt, and J.~Rawnsley, ``Some remarks on the classification of poisson lie groups,'' {\em Contemporary Mathematics} {\bfseries 179} (1994) 1--1.

\bibitem{Kawaguchi:2011pf}
I.~Kawaguchi and K.~Yoshida, ``{Hybrid classical integrability in squashed sigma models},'' \href{http://dx.doi.org/10.1016/j.physletb.2011.09.117}{{\em Phys. Lett. B} {\bfseries 705} (2011) 251--254}, \href{http://arxiv.org/abs/1107.3662}{{\ttfamily arXiv:1107.3662 [hep-th]}}.

\bibitem{Chu:1994hm}
M.-f. Chu and P.~Goddard, ``{Quantization of a particle moving on a group manifold},'' \href{http://dx.doi.org/10.1016/0370-2693(94)90977-6}{{\em Phys. Lett. B} {\bfseries 337} (1994) 285--293}, \href{http://arxiv.org/abs/hep-th/9407116}{{\ttfamily arXiv:hep-th/9407116}}.

\bibitem{Schaller:1994es}
P.~Schaller and T.~Strobl, ``{Poisson structure induced (topological) field theories},'' \href{http://dx.doi.org/10.1142/S0217732394002951}{{\em Mod. Phys. Lett. A} {\bfseries 9} (1994) 3129--3136}, \href{http://arxiv.org/abs/hep-th/9405110}{{\ttfamily arXiv:hep-th/9405110}}.

\bibitem{Ikeda:1993aj}
N.~Ikeda and K.~I. Izawa, ``{General form of dilaton gravity and nonlinear gauge theory},'' \href{http://dx.doi.org/10.1143/PTP.90.237}{{\em Prog. Theor. Phys.} {\bfseries 90} (1993) 237--246}, \href{http://arxiv.org/abs/hep-th/9304012}{{\ttfamily arXiv:hep-th/9304012}}.

\bibitem{Ikeda:1993fh}
N.~Ikeda, ``{Two-dimensional gravity and nonlinear gauge theory},'' \href{http://dx.doi.org/10.1006/aphy.1994.1104}{{\em Annals Phys.} {\bfseries 235} (1994) 435--464}, \href{http://arxiv.org/abs/hep-th/9312059}{{\ttfamily arXiv:hep-th/9312059}}.

\bibitem{Cattaneo:2001bp}
A.~S. Cattaneo and G.~Felder, ``{Poisson sigma models and deformation quantization},'' \href{http://dx.doi.org/10.1142/S0217732301003255}{{\em Mod. Phys. Lett. A} {\bfseries 16} (2001) 179--190}, \href{http://arxiv.org/abs/hep-th/0102208}{{\ttfamily arXiv:hep-th/0102208}}.

\bibitem{Grumiller:2003ad}
D.~Grumiller and W.~Kummer, ``{The Classical solutions of the dimensionally reduced gravitational Chern-Simons theory},'' \href{http://dx.doi.org/10.1016/S0003-4916(03)00138-6}{{\em Annals Phys.} {\bfseries 308} (2003) 211--221}, \href{http://arxiv.org/abs/hep-th/0306036}{{\ttfamily arXiv:hep-th/0306036}}.

\bibitem{Couvreur_2017}
R.~Couvreur, J.~L. Jacobsen, and H.~Saleur, ``Entanglement in nonunitary quantum critical spin chains,'' \href{http://dx.doi.org/10.1103/PhysRevLett.119.040601}{{\em Physical Review Letters} {\bfseries 119} no.~4, (July, 2017) }.

\bibitem{Bonderson:2017osr}
P.~Bonderson, C.~Knapp, and K.~Patel, ``{Anyonic Entanglement and Topological Entanglement Entropy},'' \href{http://dx.doi.org/10.1016/j.aop.2017.07.018}{{\em Annals Phys.} {\bfseries 385} (2017) 399--468}, \href{http://arxiv.org/abs/1706.09420}{{\ttfamily arXiv:1706.09420 [quant-ph]}}.

\bibitem{Wong:2022eiu}
G.~Wong, ``{A note on the bulk interpretation of the quantum extremal surface formula},'' \href{http://dx.doi.org/10.1007/JHEP04(2024)024}{{\em JHEP} {\bfseries 04} (2024) 024}, \href{http://arxiv.org/abs/2212.03193}{{\ttfamily arXiv:2212.03193 [hep-th]}}.

\bibitem{Verlinde:1989ua}
H.~L. Verlinde, ``{Conformal Field Theory, 2-$D$ Quantum Gravity and Quantization of Teichmuller Space},'' \href{http://dx.doi.org/10.1016/0550-3213(90)90510-K}{{\em Nucl. Phys. B} {\bfseries 337} (1990) 652--680}.

\bibitem{EllegaardAndersen:2011vps}
J.~Ellegaard~Andersen and R.~Kashaev, ``{A TQFT from Quantum Teichm{\"u}ller Theory},'' \href{http://dx.doi.org/10.1007/s00220-014-2073-2}{{\em Commun. Math. Phys.} {\bfseries 330} (2014) 887--934}, \href{http://arxiv.org/abs/1109.6295}{{\ttfamily arXiv:1109.6295 [math.QA]}}.

\bibitem{Collier:2023fwi}
S.~Collier, L.~Eberhardt, and M.~Zhang, ``{Solving 3d gravity with Virasoro TQFT},'' \href{http://dx.doi.org/10.21468/SciPostPhys.15.4.151}{{\em SciPost Phys.} {\bfseries 15} no.~4, (2023) 151}, \href{http://arxiv.org/abs/2304.13650}{{\ttfamily arXiv:2304.13650 [hep-th]}}.

\bibitem{Faddeev:1999fe}
L.~D. Faddeev, ``{Modular double of quantum group},'' in {\em {Conference Moshe Flato}}, pp.~149--156.
\newblock 2000.
\newblock \href{http://arxiv.org/abs/math/9912078}{{\ttfamily arXiv:math/9912078}}.

\bibitem{Ponsot:1999uf}
B.~Ponsot and J.~Teschner, ``{Liouville bootstrap via harmonic analysis on a noncompact quantum group},'' \href{http://arxiv.org/abs/hep-th/9911110}{{\ttfamily arXiv:hep-th/9911110}}.

\bibitem{Ponsot:2000mt}
B.~Ponsot and J.~Teschner, ``{Clebsch-Gordan and Racah-Wigner coefficients for a continuous series of representations of U(q)(sl(2,R))},'' \href{http://dx.doi.org/10.1007/PL00005590}{{\em Commun. Math. Phys.} {\bfseries 224} (2001) 613--655}, \href{http://arxiv.org/abs/math/0007097}{{\ttfamily arXiv:math/0007097}}.

\bibitem{Kharchev:2001rs}
S.~Kharchev, D.~Lebedev, and M.~Semenov-Tian-Shansky, ``{Unitary representations of $U_q (\mathfrak{sl}(2, \mathbb{R}))$, the modular double, and the multiparticle q deformed Toda chains},'' \href{http://dx.doi.org/10.1007/s002200100592}{{\em Commun. Math. Phys.} {\bfseries 225} (2002) 573--609}, \href{http://arxiv.org/abs/hep-th/0102180}{{\ttfamily arXiv:hep-th/0102180}}.

\bibitem{Bytsko:2002br}
A.~G. Bytsko and J.~Teschner, ``{R operator, coproduct and Haar measure for the modular double of U(q)(sl(2,R))},'' \href{http://dx.doi.org/10.1007/s00220-003-0894-5}{{\em Commun. Math. Phys.} {\bfseries 240} (2003) 171--196}, \href{http://arxiv.org/abs/math/0208191}{{\ttfamily arXiv:math/0208191}}.

\bibitem{Bytsko:2006ut}
A.~G. Bytsko and J.~Teschner, ``{Quantization of models with non-compact quantum group symmetry: Modular XXZ magnet and lattice sinh-Gordon model},'' \href{http://dx.doi.org/10.1088/0305-4470/39/41/S11}{{\em J. Phys. A} {\bfseries 39} (2006) 12927--12981}, \href{http://arxiv.org/abs/hep-th/0602093}{{\ttfamily arXiv:hep-th/0602093}}.

\bibitem{Atiyah:1982fa}
M.~F. Atiyah and R.~Bott, ``{The Yang-Mills equations over Riemann surfaces},'' {\em Phil. Trans. Roy. Soc. Lond. A} {\bfseries 308} (1982) 523--615.

\bibitem{Alekseev:1993rj}
A.~Y. Alekseev and A.~Z. Malkin, ``{Symplectic structure of the moduli space of flat connection on a Riemann surface},'' \href{http://dx.doi.org/10.1007/BF02101598}{{\em Commun. Math. Phys.} {\bfseries 169} (1995) 99--120}, \href{http://arxiv.org/abs/hep-th/9312004}{{\ttfamily arXiv:hep-th/9312004}}.

\bibitem{weinstein1990affine}
A.~Weinstein, ``Affine poisson structures,'' {\em Internat. J. Math} {\bfseries 1} no.~3, (1990) 343--360.

\end{thebibliography}\endgroup
\end{CJK*}
\end{document}